\documentclass[acmtog, dvipsnames, screen=true, authorversion]{acmart} %

\usepackage{tabularx,booktabs} %
\usepackage{hyphenat,balance}
\usepackage[fleqn,tbtags]{mathtools}
\usepackage{overpic}  %
\usepackage{wrapfig}
\usepackage{contour}
\usepackage{microtype}
\usepackage{multirow}
\usepackage{colortbl}
\usepackage{amsmath}
\usepackage{bbm}
\usepackage{textcomp}
\usepackage{enumitem}
\setitemize{noitemsep,topsep=0pt,parsep=0pt,partopsep=0pt}
\usepackage{derivative}
\usepackage{float}

\usepackage{graphicx}
\graphicspath{{./images/}}

\definecolor{mypink1}{RGB}{205,115,114}

\usepackage[linesnumbered,ruled,lined]{algorithm2e} 
\SetAlFnt{\small}
\SetAlCapFnt{\small}
\SetAlCapNameFnt{\small}
\SetAlCapHSkip{0pt}

\usepackage{epigraph}
\usepackage{algpseudocode}
\usepackage{mathrsfs}
\usepackage{makecell}

\usepackage{amsmath,amsfonts,bm}

\def\1{\bm{1}}

\def\va{{\bm{a}}}

\def\vc{{\bm{c}}}

\def\vr{{\bm{r}}}

\def\vv{{\bm{v}}}

\def\vx{{\bm{x}}}

\def\vz{{\bm{z}}}

\DeclareMathAlphabet{\mathsfit}{\encodingdefault}{\sfdefault}{m}{sl}
\SetMathAlphabet{\mathsfit}{bold}{\encodingdefault}{\sfdefault}{bx}{n}

\newcommand{\vepsilon}{\mathbf{\epsilon}}

\def \etal {{\emph{et al}.\thinspace}}
\def \eg {{\emph{e.g.},\thinspace}}
\def \ie {{\emph{i.e.},\thinspace}}

\newcommand{\rev}[1]{{\color{black} #1}}
\newcommand{\mypar}[1]{\vspace{1mm}\noindent\textbf{#1}}

\newcommand{\name}{\emph{Sketch2Anim}\xspace}

\setcopyright{cc}
\setcctype{by-nc}
\acmJournal{TOG}
\acmYear{2025} \acmVolume{44} \acmNumber{4} \acmArticle{} \acmMonth{8} \acmPrice{}\acmDOI{10.1145/3731167}

\begin{document}

\title{\name: Towards Transferring Sketch Storyboards into 3D Animation}

\author{Lei Zhong}
\orcid{0000-0003-1778-9282}
\affiliation{%
 \institution{University of Edinburgh}
 \streetaddress{10 Crichton Street}
 \city{Edinburgh}
 \country{United Kingdom}
 }
\email{zhongleilz@icloud.com}

\author{Chuan Guo}
\orcid{0000-0002-4539-0634}
\affiliation{%
 \institution{Snap Inc.}
 \streetaddress{New York City}
 \city{New York}
 \country{United States}
}
\email{guochuan5513@gmail.com}

\author{Yiming Xie}
\orcid{0000-0002-2408-403X}
\affiliation{%
 \institution{Northeastern University}
 \streetaddress{360 Huntington Ave}
 \city{Boston}
 \country{United States}
}
\email{ymxyimingxie@gmail.com}

\author{Jiawei Wang}
\orcid{0009-0009-7343-8066}
\affiliation{%
 \institution{University of Edinburgh}
 \streetaddress{10 Crichton Street}
 \city{Edinburgh}
 \country{United Kingdom}
 }
 \email{jiaweiwang0222@gmail.com}

\author{Changjian Li}
\orcid{0000-0003-0448-4957}
\affiliation{%
 \institution{University of Edinburgh}
 \streetaddress{10 Crichton Street}
 \city{Edinburgh}
 \country{United Kingdom}
 }
 \email{chjili2011@gmail.com}

\begin{teaserfigure}
    \centering
    \begin{overpic}[width=\textwidth]{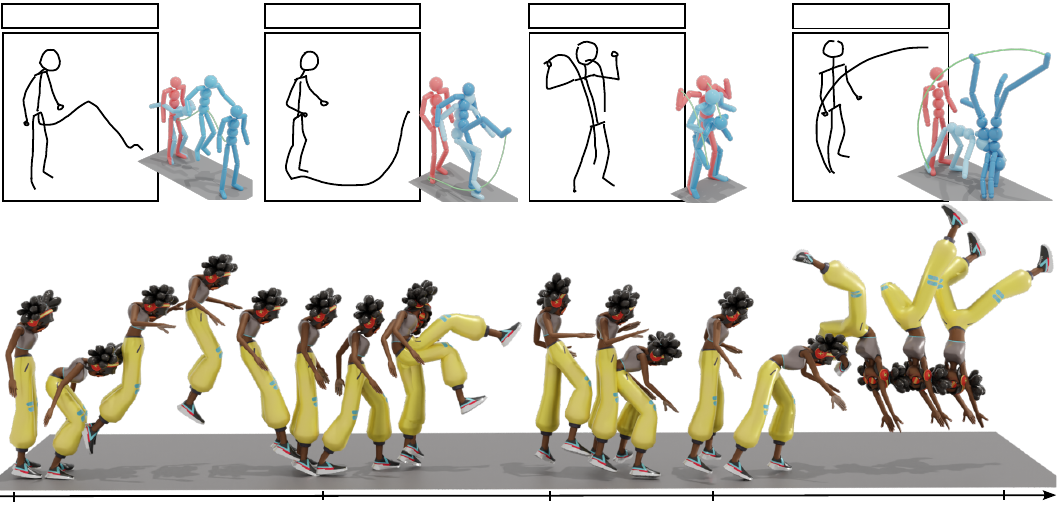} %
        \put(14,0.6) {\small \textbf{Jump}}
        \put(40,0.6)  {\small \textbf{Kick}}
        \put(58,0.6) {\small \textbf{Bow}}
        \put(78.5,0.6) {\small \textbf{Backflip}}
        \put(96.2,0.2) {\small \textbf{Time}}
        \put(0.7,47.2) {\small \textbf{Action: Jump}}
        \put(25.6,47.2) {\small \textbf{Action: Kick}}
        \put(50.7,47.2) {\small \textbf{Action: Bow}}
        \put(75.7,47.2) {\small \textbf{Action: Backflip}}
    \end{overpic}
    \vspace{-7mm}
    \caption{\textbf{Storyboard animation}. Given a storyboard (top row) composed of a few sketch keyposes, joint trajectories, and the corresponding action words, we propose \name, a novel approach that automatically transfers each sketch frame into a 3D motion clip, where the keypose is highlighted in red, and the trajectory is highlighted in green. All these clips are then blended into a complete and coherent animation (bottom row) depicting the storyboard.
    }
    \label{fig:teaser}
\end{teaserfigure}
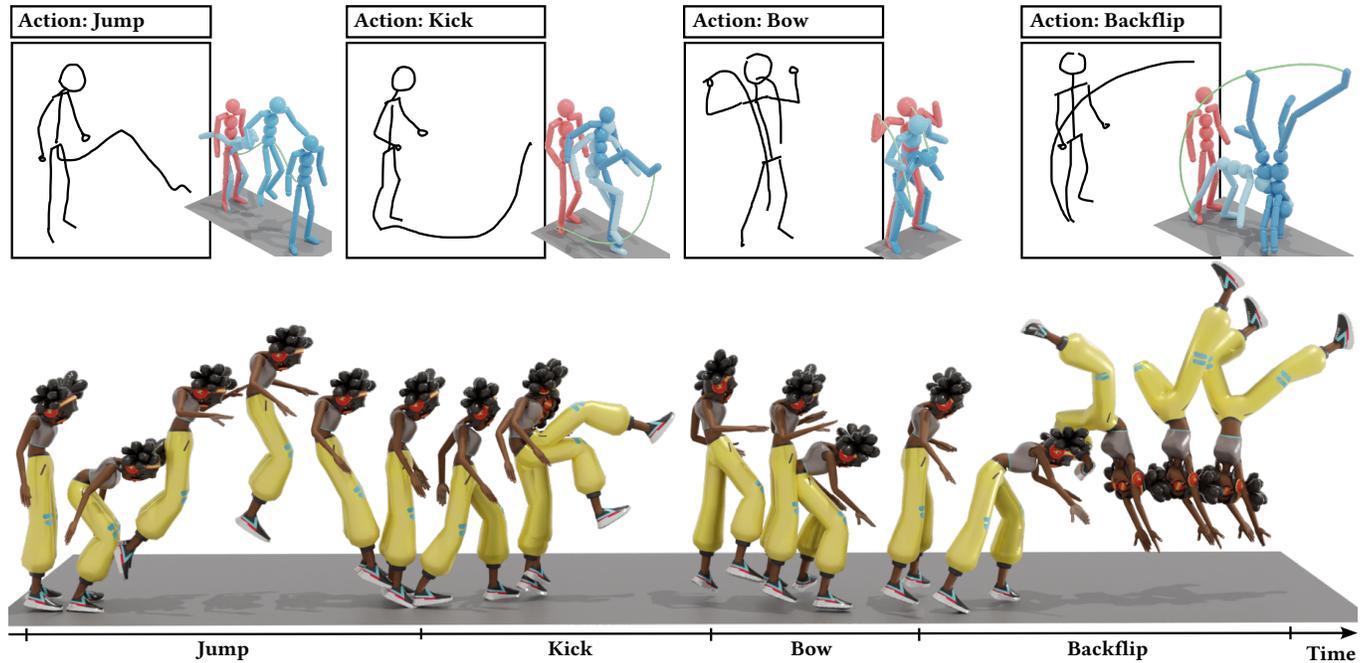

\begin{abstract}
Storyboarding is widely used for creating 3D animations. Animators use the 2D sketches in storyboards as references to craft the desired 3D animations through a trial-and-error process. The traditional approach requires exceptional expertise and is both labor-intensive and time-consuming. Consequently, there is a high demand for automated methods that can directly translate 2D storyboard sketches into 3D animations.
This task is under-explored to date and 
inspired by the significant advancements of motion diffusion models, we propose to address it
from the perspective of conditional motion synthesis.
We thus present \name, composed of two key modules for sketch constraint understanding and motion generation.
Specifically, due to the large domain gap between the 2D sketch and 3D motion, instead of directly conditioning on 2D inputs, we design a 3D conditional motion generator that simultaneously leverages 3D keyposes, joint trajectories, and action words, to achieve precise and fine-grained motion control.
Then, we invent a neural mapper dedicated to aligning user-provided 2D sketches with their corresponding 3D keyposes and trajectories in a shared embedding space, enabling, \emph{for the first time}, direct 2D control of motion generation. 
Our approach successfully transfers storyboards into high-quality 3D motions and inherently supports direct 3D animation editing, thanks to the flexibility of our multi-conditional motion generator.
Comprehensive experiments and evaluations, and a user \rev{perceptual} study demonstrate the effectiveness of our approach.

\emph{The code, data, trained models, and sketch-based motion designing interface \rev{are at  \href{https://zhongleilz.github.io/Sketch2Anim/}{https://zhongleilz.github.io/Sketch2Anim/}}}. %

\end{abstract}

\ccsdesc[500]{Computing methodologies~Animation}

\keywords{Story Board, Sketch-based Animation, Motion synthesis}

\maketitle

\section{Introduction}

\epigraph{"Animation is not the art of drawings that move but the art of movements that are drawn."}{— Norman McLaren}

\begin{figure}[!t]
    \centering
    \begin{overpic}[width=0.97\linewidth]{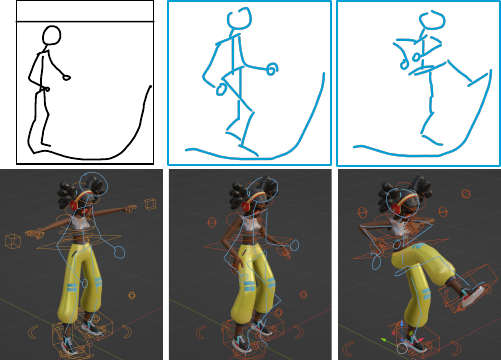} %
        \put(4.5,68.6) {\small Action: Kick}
    \end{overpic}
    \vspace{-2mm}
    \caption{\textbf{Traditional 3D animation workflow.} Given one frame of a sketch storyboard (top-left), animators first imagine the desired keypose sequence guided by the trajectory and the action word. Two such representative keyposes are shown in blue. These keyposes (including the first keypose) are imported into 3D software (\eg Blender) as keypose and motion references. The animators then manually place the pre-defined 3D joints to match the keypose, while crafting the desired motion to explain the trajectory and action word. After a long trial-and-error process, a high-quality animation is produced. \emph{See the supplemental video for the whole manual process}.
    }
    \vspace{-5mm}
    \label{fig:workflow}
\end{figure}

Sketch storyboards are widely used by artists to quickly outline their creative animation ideas \cite{lasseter1998principles,williams2012animator}. As illustrated in Figs.~\ref{fig:teaser} and \ref{fig:workflow}, a storyboard typically consists of a series of 2D sketch frames representing the desired animation clips.
Each frame features an abstracted character sketch depicting the keypose, accompanied by joint trajectories that convey the envisioned animation sequence. Additionally, a single action word is often tagged to indicate the semantic and motion space.
For instance, as shown in the second frame in Fig.~\ref{fig:teaser}, the word ``Kick'' constrains the motion space, while the keypose describes the starting pose and the trajectory describes the specific leg movement.
In the current animation workflow (Fig.~\ref{fig:workflow}), animators use these 2D sketches as references and manually align pre-defined 3D character joints to match the 2D keypose, while adding 3D joint motions to follow the 2D trajectories through a trial-and-error process. This workflow requires exceptional expertise developed over years of practice and is both labor-intensive and time-consuming.
Therefore, automated methods that can directly translate 2D storyboards into 3D animations are highly demanded.

Earlier research closely relevant to this task is sketch-based motion retrieval~\cite{thorne2004motion,yoo2014sketching}. They usually extract handcrafted features (\eg joint angles and bone vectors) from input sketches and search for the closest 3D animation in a pre-collected database.  %
However, these methods are constrained by the scope of pre-existing datasets, resulting in limited scalability and expressiveness. When dealing with multiple sketch types (\eg keyposes and trajectories), it is often hard to design proper matching score functions, leading to inferior retrieved animations.

More recently, diffusion models have demonstrated remarkable performance in motion generation~\cite{tevet2023human,chen2023executing,goel2024iterative,athanasiou2024motionfix,barquero2024seamless}.
Inspired by the great advancements, we propose to solve the translation from 2D storyboards to 3D motions from the perspective of conditional motion synthesis.
With this key idea, the translation can be reformulated as two sub-problems: i) \emph{how to obtain the accurate 2D keypose and trajectory from the raw sketch of a storyboard} and ii) \emph{how to produce a high-quality motion clip conditioned on the 2D keypose, trajectory, and action word?}
This first sub-problem is well-studied, \eg Sketch2Pose~\cite{brodt2022sketch2pose} can robustly detect 2D keyposes from sketches and the 2D trajectory can be traced precisely in the user interface. In this paper, we focus on the second sub-problem, \ie multi-conditional motion generation.

Intuitively, conditioning a motion diffusion model on 2D keyposes, joint trajectories, and action words offers a straightforward approach. However, due to our unique problem setting, the considerable domain gap between 2D inputs and 3D motion makes this solution infeasible. Specifically, the detected 2D keypose and joint trajectories are the projection of the imagined 3D motion. The former is in the pixel space, while the latter is in the 4D space (\ie space and time). High-quality motions directly from 2D conditional inputs are hard to learn (see Sec.~\ref{subsec:abl_study}).
In contrast, 3D keyposes and trajectories are more informative, providing concrete guidance for motion generation.
However, even with well-defined 3D keypose and trajectory, the previous conditional motion generation methods are incapable of consuming all of them, since they are typically limited to handling one or two control signals, such as text or text paired with 3D trajectories~\cite{xie2024omnicontrol}, 3D indoor scene~\cite{yi2024tesmo}, or 3D contacting objects~\cite{diller2024cg,li2025controllable}. Controlling models to simultaneously adhere to more than two conditions, such as action words, keyposes, and trajectories remains a significant challenge.

To address the above challenges, we propose a novel approach, dubbed \name, composed of two core modules. Our design is inspired by the two key operators in the traditional animation workflow - lifting the 2D keypose and trajectory to 3D by 2D-3D alignment and crafting the vivid motion based on the lifted constraints. 
Specifically, instead of mathematically lifting the 2D keypose and trajectory into 3D, we design a neural mapper (Sec.~\ref{sec:mapper}) dedicated to aligning 2D keypose and trajectory to their corresponding 3D keypose and trajectory in the shared embedding space.
This shared embedding space enables seamless acceptance of 2D keyposes and joint trajectories as input during inference, and more informative and precise 3D keypose and trajectory as conditions during training.

Our second module is a multi-conditional motion generator (Sec. \ref{sec:generator}), where we successfully inject both the keypose and trajectory controls on top of the action word.
Although the 3D keypose and trajectory can be combined through joint positions and treated as a single control signal, they actually emphasize different motion properties.
The keypose primarily imposes local static constraints at specific timesteps, while a joint trajectory conveys both global and local dynamic movements. 
To better balance their respective influences during training, we propose a trajectory ControlNet combined with a trajectory-aware keypose adapter.
The mechanism is that the adapter that is positioned between the trajectory ControlNet and the motion diffusion model, refines the output features from the trajectory ControlNet using the keypose condition and integrates the refined residual features into the diffusion model.
This design enables the ControlNet to focus on dynamic motion control, while the adapter refines the local pose at a specific timestep,
effectively minimizing interference between the two conditions.
We finally blend the animation clips from each keyframe into a complete and coherent 3D animation explaining the storyboard. 
With the new approach, we are able to translate the 2D sketch storyboard directly into its 3D animation, and in the meantime, it naturally supports 3D motion editing, benefiting from our motion generator.

We evaluate our method on a few synthetic benchmarks and real user sketches, and in both cases, \name successfully transfers sketch storyboards into high-quality animations. Extensive experiments and ablation studies validate the effectiveness of our method. 

In summary, the principal contributions of this work include:
\begin{itemize}
    \item We introduce a neural mapper to align the embedding space of 2D and 3D keyposes and trajectories. This shared embedding space allows us to leverage 3D control signals as 2D surrogates during training and then seamlessly use 2D control signals during inference.

    \item We present a keypose adapter integrated with the trajectory ControlNet, simultaneously controlling the motion diffusion model to effectively adhere to both keypose and joint trajectory constraints.

    \item Combined together, these technical contributions enable the \emph{first} approach that adapts the motion diffusion model to generate 3D animations directly from 2D storyboards, and an accompanying sketch-based animation design system.
    
\end{itemize}

\section{Related Work}
In this section, we summarize prior works from sketch-based 3D character animation and human motion synthesis that are closely related to our task. %

\paragraph{Sketch-based 3D Character Animation}
Sketch-based 3D character animation provides intuitive tools for artists to create lifelike animations directly from sketches~\cite{choi2016sketchimo,peng2021sketch,zhou2024drawingspinup}.
Before attending to 3D animation, early works study the problem of inferring static 3D poses from various types of sketches, such as 2D stick figures~\rev{\cite{lin2010sketching,hahn2015sketch,davis2006sketching}}, lines of action~\cite{guay2013line}, custom sketches~\cite{brodt2022sketch2pose}, and silhouette sketches~\cite{bessmeltsev2016gesture3d}.
The estimated 3D pose is then used to pose a pre-defined 3D character.
For instance, The Line of Action~\cite{guay2013line} introduces a sketch-based interface that allows users to draw a single aesthetic line of action to pose an existing 3D character by solving an optimization.
Lin \etal \shortcite{lin2010sketching} propose to use stick figure sketches for sitting pose design, while Hahn \etal \shortcite{hahn2015sketch} extend the usability to general characters with their rough stick-figure sketch. 
More recently, Brodt \etal \shortcite{brodt2022sketch2pose} propose Sketch2Pose, which can handle more complicated sketches. Their method first predicts 2D joints and body part contacts, and optimizes the 3D pose of an SMPL model to conform to the sketch. Based on our experiments, their 2D joint detection in our storyboard scenario is robust, but their 3D keypose is less satisfactory. Thus, as a pre-processing step, we use their method to obtain the 2D joints for our approach. 

Other than 3D poses, as a step forward, some works start exploring the dynamic motion modeling from 2D sketches, \eg simple 2D paths~\cite{igarashi1998path,thorne2004motion} and space-time curves~\cite{guay2015space}.
MotionDoodles~\cite{thorne2004motion} presents a sketch-based interface that allows users to create character motions by drawing gesture sketches. They build a gesture vocabulary to drive the character to perform desired motions. 
Yoo \etal \shortcite{yoo2014sketching} propose a pipeline for generating 3D human character motions by retrieving motion sequences from a database. Their input sketch consists of character strokes, motion curves, and rotation curves, and is analogous to ours, especially since the latter two curves are similar to trajectory strokes.
However, these methods are retrieval-based and depend highly on predefined motion templates or databases, limiting their adaptability to complex or unseen motions. In contrast, our system overcomes this limitation by synthesizing novel motions through a motion diffusion model.

Closest to our work, the concurrent study DoodleYourMotion~\cite{wu2024doodle} proposes a sketch-conditioned motion generation method. Their approach utilizes a silhouette sketch image as a control signal to guide the motion diffusion model and introduces a sketch-aware local attention mechanism to enhance interactions between local sketch patches and motion keyframes. However, by relying solely on the sketch image as the control signal, their method is constrained to influencing only the keyframes in the generated motion, lacking the ability to effectively control the motion's dynamics. 
As a comparison, our sketch system exploits more simple stick figure style character sketches to ease user burden and supports trajectory strokes to enable finer-grained motion control. 

In parallel, another line of research aims to animate static images or drawings~\cite{weng2019photo,smith2023method,yang2022object,zhou2024drawingspinup,dvorovzvnak2020monster}.
They first model a textured 3D mesh from the input and further model the 3D motion either by rigging the character to the pre-defined animation-ready 3D assets or by manually dragging the mesh parts in their interface.
However, their main focus is 3D character modeling, while we concentrate on creating high-quality 3D animations from sketch storyboards by understanding the motion cues in the input sketches.

\begin{figure*}[!t]
    \centering
    \begin{overpic}[width=\textwidth]{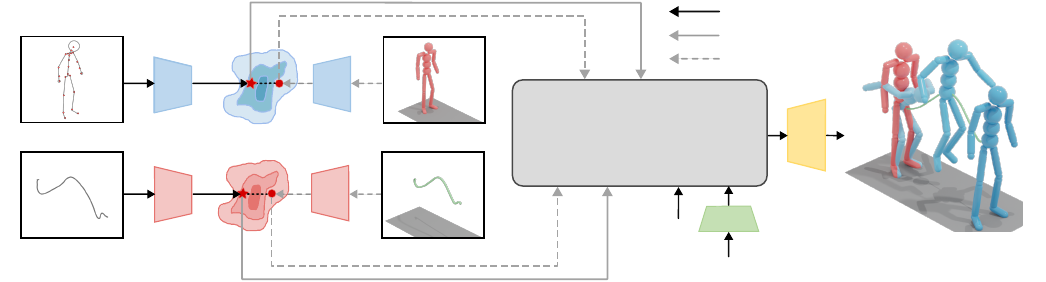} %
        \put(3,13.5) {\small 2D keypose}
        \put(2.7,2.5) {\small 2D trajectory}
        \put(11.9,19.6) {\small $\mathbf{K}_{2D}$}
        \put(11.9,9.5) {\small $\mathbf{T}^r_{2D}$}
        \put(15,18.5) {\small $\mathcal{E}_{k}^{2D}$}
        \put(15,8) {\small $\mathcal{E}_{tr}^{2D}$}
        \put(30,18.5) {\small $\mathcal{E}_{k}^{3D}$}
        \put(30,8) {\small $\mathcal{E}_{tr}^{3D}$}
        \put(19,13.7) {2D-3D Mapper}
        \put(21.5,12) {\textcolor{MidnightBlue}{\emph (Sec.~\ref{sec:mapper})}}
        \put(33.7,19.6) {\small $\mathbf{K}_{3D}$}
        \put(33.6,9.5) {\small $\mathbf{T}^r_{3D}$}
        \put(37.5,13.5) {\small 3D keypose}
        \put(37.2,2.5) {\small 3D trajectory}
        \put(54,15.5) {\bf Multi-conditional}
        \put(54,13) {\bf Motion Generator}
        \put(58,11) {\textcolor{MidnightBlue}{\emph (Sec.~\ref{sec:generator})}}
        \put(61,7) {\footnotesize Noisy}
        \put(61,5.5) {\footnotesize latent}
        \put(65.2,6.5) {\small $\vz_t$}
        \put(66.2,0.5) {\small $\mathbf{W}_a$=``Jump''}
        \put(68.1,5.4) {\small CLIP}
        \put(70.3,7.7) {\small $\mathbf{a}$}
        \put(76.2,13.7) {\small $\mathcal{D}$}
        \put(83,2) {\bf Generated Motion}
        \put(69.5, 25.3) {\small Training and Inference}
        \put(69.5, 23) {\small Inference}
        \put(69.5, 20.8) {\small Training}
    \end{overpic}
    \vspace{-6mm}
    \caption{\textbf{Overview of \name.} Our pipeline consists of two core modules - the multi-conditional motion generator (Sec.~\ref{sec:generator}) and the 2D-3D neural mapper (Sec.~\ref{sec:mapper}). Instead of directly lifting the 2D keypose and trajectory into their 3D counterparts, we train a neural mapper dedicated to aligning the two domains in the embedding space. Because of this shared embedding, it enables the employment of more informative and precise 3D keyposes and trajectories as the motion conditions in the motion generator, while exploiting the 2D keypose and trajectory detected from the sketch storyboard at inference time. The legend indicates the data flow at training and inference of both modules\rev{. See} the following sections for detailed technical designs.
    }
    \label{fig:pipeline}
\end{figure*}
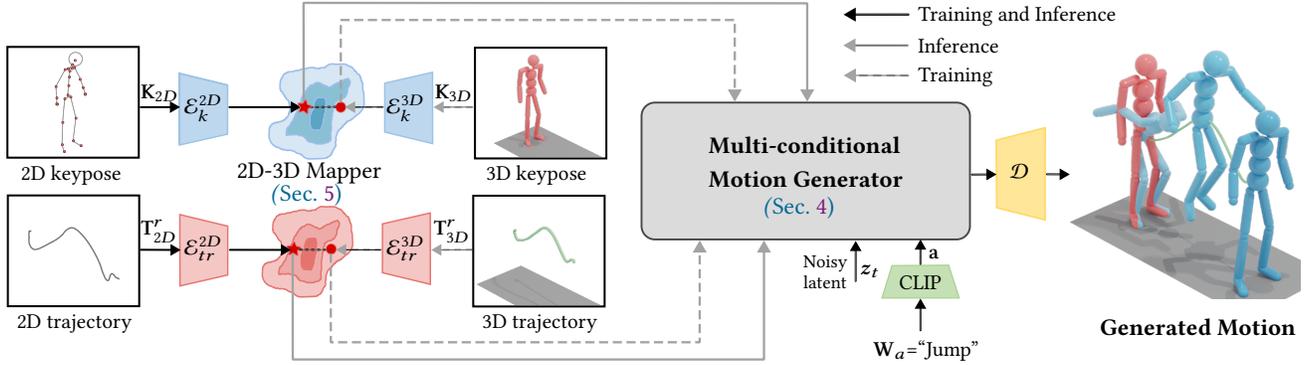

\paragraph{Human Motion Synthesis}
Human motion synthesis focuses on generating diverse and realistic human-like movements.
Traditional approaches rely on concatenation and retrieval techniques, including motion graphs~\cite{kovar2023motion,heck2007parametric,shin2006fat} and motion matching~\cite{buttner2015motionmatching}.
Recently, learning-based methods have been widely adopted to generate high-quality and realistic motions, utilizing models such as GANs~\cite{li2022ganimator,ghosh2021synthesis}, autoregressive GPTs~\cite{guo2022tm2t,jiang2023motiongpt,zhang2023generating,zhangmotiongpt}, generative masked Transfomers~\cite{pinyoanuntapong2024mmm,guo2024momask,li2024lamp}, and diffusion models~\cite{tevet2023human,motiondiffuse,chen2023executing,zhou2025emdm}.

Driven by the efficacy of diffusion models for conditioning, several works utilize diffusion models to enable motion in-betweening~\cite{cohan2024flexible,agrawal2024skel}, human-object interactions~\cite{li2023object,li2025controllable}, and trajectories control~\cite{xie2024omnicontrol,karunratanakul2023gmd,dai2025motionlcm,wan2023tlcontrol}.
To harness the capability of pre-trained motion diffusion models, OmniControl~\cite{xie2024omnicontrol} employs a ControlNet~\cite{zhang2023adding} to guide these models in generating motions that adhere to 3D joint trajectories. 
Similarly, SMooDi~\cite{zhong2025smoodi} employs ControlNet to steer these models 
for stylized motion generation.
However, both methods utilize ControlNet for a single condition. When handling multiple conditions, how to effectively ensemble multiple ControlNets remains an open problem, even in the image and video domains.
In this work, we propose a trajectory-aware keypose adapter designed to work alongside the trajectory ControlNet, guiding the pre-trained diffusion model to accommodate multiple conditions.%

In terms of multiple conditions, SKEL-Betweener~\cite{agrawal2024skel} is recently proposed and able to accept both the 3D keypose and 3D trajectory as conditions in its motion in-betweening task to synthesize novel motions using their skeletal transformer.
It is worth noting that their start and end keyposes are in the 3D format and well-placed spatially with correct orientations. 
However, in our unique setting, it is difficult to obtain the 3D keypose and trajectory from 2D sketches, and impossible to derive the global placement information from the individually drawn storyboard frames, which differentiates us from them, making their techniques inapplicable.

\begin{figure}[!t]
    \centering
    \begin{overpic}[width=0.95\linewidth]{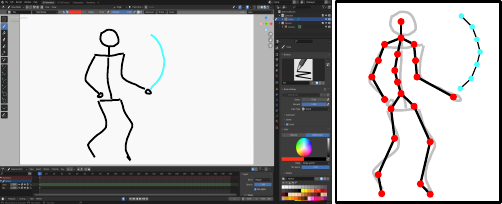} %
        \put(4.5,35) {\small (a)}
        \put(34,34) {\footnotesize \textbf{HandLifting}}
        \put(95,37) {\small (b)}
    \end{overpic}
    \vspace{-2mm}
    \caption{\textbf{User interface and joint detection.} (a) In the interface, the white canvas is the main drawing area. Users can sketch or type the action word. We use different pens for character strokes and trajectories. %
    (b) We use Sketch2Pose to detect the 2D joint points (red points), which serve as our input, instead of the raw sketch in (a); and we draw the line segments between joint points only for intuitive visualization. The uniformly re-sampled trajectory points are drawn in cyan.
    }
    \vspace{-4mm}
    \label{fig:user_interface}
\end{figure}

\section{Overview}
\label{sec:overview}
Our novel approach produces high-quality 3D animations directly from a 2D sketch storyboard. 

\mypar{Storyboard}. 
Figs. \ref{fig:teaser} and \ref{fig:workflow} display examples of 2D storyboards. To be clear, in our setting, for each frame, we only consider \emph{character strokes} indicating the static pose, \emph{joint trajectory strokes} (if any) expressing dynamic movements, and the \emph{action word} describing the motion content, excluding decorative sketches depicting the scene. 

We have developed a user interface as a Blender add-on (see Fig.~\ref{fig:user_interface}(a)). Character strokes in black and joint trajectories in cyan are traced separately in a \emph{known} camera view. Users can either sketch the action word or type the text. For simplicity, the character is drawn in the stick-figure style with circles and line segments indicating the head, body, and limbs. Storyboard frames are drawn sequentially from the $3/4$ view without global placements, and users finally execute \name to obtain the complete animation.

\mypar{Input}.
Suppose there are $f$ sketches in the storyboard, %
and for each of them, the character strokes are rendered into a rasterized image.
The image is first fed into Sketch2Pose \cite{brodt2022sketch2pose} to obtain a set of 2D joint points. Following the joint setting in the SMPL body \cite{loper2023smpl}, we simply augmented the point set to obtain the input 2D joints indicating the static keypose $\{(x_j, y_j)\}_1^{J=22}$, where $J$ is the number of joints. 
As for the joint trajectory, the user is able to draw multiple strokes for multiple joints. We trace and downsample each stroke, and attach it to the corresponding joint by closest points checking, resulting in a set of 2D trajectory points $\{(x_i^j, y_i^j)\}_{i=1}^{t_r}$. Note that, the superscript $j$ represents the joint index, and $t_r$ varies based on the length of the trajectory stroke.
An example of joints and trajectory points is shown in Fig.~\ref{fig:user_interface}(b). 
Lastly, the action word $\mathbf{W}_a$ is pure text input. If users sketch the word, we use an OCR \cite{bautista2022scene} approach to detect it. %

\mypar{Output}.
Given the 2D joint points, 2D trajectory points, and the action word from a single sketch frame, the algorithm generates a 3D motion clip $\bm{m}$. By default, it consists of $N=40$ frames (\ie $2s$) in our setting.
A final 3D motion by blending the $f$ motion clips is produced, denoted as $\mathcal{M}$. %
For visualization purposes, we occasionally employ special characters (\ie `Michelle' and `Clair' from Mixamo characters in Figs. \ref{fig:teaser} and \ref{fig:editing}, respectively) by retargeting \cite{keemap_blender_2024}. Unless otherwise stated, a default \rev{capsule-bone human model} is exploited throughout the paper.

\mypar{Feature representation}.
We adopt a unified matrix representation for the joint (\ie keypose), trajectory, and motion. 
Specifically, the 2D keypose (22 2D joints) is filled into the \emph{first} or \emph{last} motion frame, denoted as $\mathbf{K}_{2D} \in \mathbb{R}^{N \times J \times 2}$, and its 3D counterpart is represented as $\mathbf{K}_{3D} \in \mathbb{R}^{N \times J \times 3}$. %
Similarly, joint positions along successive frames represent the trajectory of a specific joint. 
The trajectory points for the joint $j$ are filled into the \emph{first} or \emph{last} $t_r$ motion frames of the $j$-th joint, denoted as $\mathbf{T}^r_{2D} \in \mathbb{R}^{N \times J \times 2}$, while the 3D counterpart is denoted as $\mathbf{T}^r_{3D} \in \mathbb{R}^{N \times J \times 3}$. Note that, $\mathbf{T}^r_{2D}$ and $\mathbf{T}^r_{3D}$ can contain more than one \rev{trajectory}.
To be more clear, if the trajectory stroke starts from its respective joint, the keypose condition is placed at the $1^{st}$/40 frame of the motion clip, while the trajectory is placed at the first $t_r$/40 frames of the motion clip. Otherwise, if the trajectory stroke ends at its respective joint, the keypose condition is at the $40^{th}$/40 frame and the trajectory is placed at the last $t_r$/40 frames.

Following HumanML3D~\cite{Guo_2022_CVPR}, we adopt a redundant motion representation that includes both local and global positions and rotations. The representation is also in a unified matrix format. Thus, the motion clip $\bm{m}$ and the final motion $\mathcal{M}$ is in the space of $\mathbb{R}^{N \times D}$ and $\mathbb{R}^{f \times N \times D}$, respectively. $D=263$ is the total dimension of the joint-based motion representation.

\mypar{The algorithm}.
An overview of our algorithm composed of two core modules is displayed in Fig. \ref{fig:pipeline}. 
The first module (Sec. \ref{sec:generator}) is a multi-conditional motion generator, which is built upon the motion diffusion model~\cite{chen2023executing}. 
It performs diffusion and denoising steps within a pre-trained latent space. During training, we use the 3D joint points (\ie the keypose) and 3D trajectory points as the surrogates for the 2D inputs.
Particularly, the designed trajectory ControlNet and the keypose adapter are integrated into the motion diffusion model, enabling robust and precise multiple motion control.
The second module (Sec. \ref{sec:mapper}) exploits a pair of encoders to align the 2D and 3D keypose and trajectory in the shared embedding space, bridging two domains.
In our implementation, the algorithm runs sequentially for each sketch frame, and the resulting motion clip is finally blended (Sec. \ref{sec:blending}).

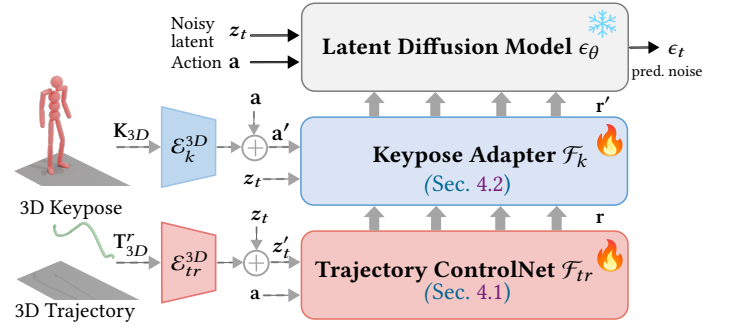
\begin{figure}[!t]
    \centering
    \begin{overpic}[width=1.12\linewidth]{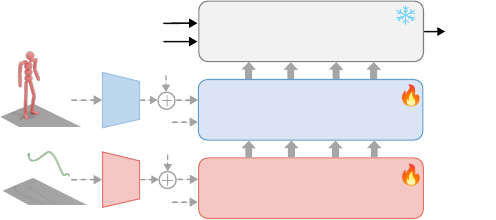} %
        \put(1,14.7) {\small 3D Keypose}
        \put(0.3,0.5) {\small 3D Trajectory}
        \put(14,25.5) {\small $\mathbf{K}_{3D}$}
        \put(14,10) {\small $\mathbf{T}^r_{3D}$}
        \put(21.8,23.2) {\small $\mathcal{E}_{k}^{3D}$}
        \put(21.8,7) {\small $\mathcal{E}_{tr}^{3D}$}
        \put(22,40.5) {\footnotesize Noisy}
        \put(22,38) {\footnotesize latent}
        \put(30,39.5) {\small $\vz_t$}
        \put(22,35) {\footnotesize Action}
        \put(30,35) {$\mathbf{a}$}
        \put(33,30) {$\mathbf{a}$}
        \put(36,25) {$\mathbf{a}'$}
        \put(32,19) {\small $\vz_t$}
        \put(33,14) {\small $\vz_t$}
        \put(36,9.5) {\small $\vz_t'$}
        \put(33,3) {\small $\mathbf{a}$}
        \put(43,37) {\bf Latent Diffusion Model $\epsilon_\theta$}
        \put(50,22) {\bf Keypose Adapter $\mathcal{F}_k$}
        \put(57,18) {\textcolor{MidnightBlue}{\emph (Sec.~\ref{subsec:adaptor})}}
        \put(42,6) {\bf Trajectory ControlNet $\mathcal{F}_{tr}$}
        \put(57,3) {\textcolor{MidnightBlue}{\emph (Sec.~\ref{subsec:trajectory_control})}}
        \put(81,13.5) {\small $\mathbf{r}$}
        \put(81,29) {\small $\mathbf{r}'$}
        \put(91,37) {\small $\epsilon_{t}$}
        \put(86,34) {\scriptsize pred. noise}
    \end{overpic}
    \caption{\textbf{The structure of our motion generator.} A trajectory ControlNet is exploited to incorporate the trajectory control, while a keypose adapter is placed between the ControlNet and the diffusion model to inject the keypose condition. 
    \rev{The same color and style of the common elements and data flow as in Fig. \ref{fig:pipeline} are used.}
    Refer to Sec.~\ref{sec:generator} for a detailed explanation.}
    \vspace{-3mm}
    \label{fig:generator}
\end{figure}

\section{Multi-conditional Motion Generator}
\label{sec:generator}

The core of our system is a multi-conditional motion latent diffusion model in which the diffusion process operates within the latent space. Fig.~\ref{fig:generator} displays an overview of this module.

Let $\vepsilon_\theta$ denote the latent denoiser (a UNet parameterized by $\theta$), and $\{\vz_t\}_{t=0}^T$ denote the sequence of noisy latents, where $\vz_T$ is a Gaussian noise.
Given a 3D keypose $\mathbf{K}_{3D}$, joint trajectories $\mathbf{T}^r_{3D}$, and an action word embedding $\mathbf{a}$ (obtained from $\mathbf{W}_a$ via CLIP~\cite{radford2021learning}), the denoising process at step $t~(0 < t \leq T)$ is defined as $\epsilon_{t} = \epsilon_\theta (\vz_t, t, \mathbf{a}, \mathbf{K}_{3D}, \mathbf{T}^r_{3D})$. A cleaner noisy latent $\vz_{t-1}$ is obtained by subtracting $\epsilon_{t}$ from $\vz_t$.
This denoising step is repeated for $T$ iterations until a clean latent $\vz_0$ is generated.
Finally, $\vz_0$ is decoded by a motion decoder $\mathcal{D}$ into a realistic motion sequence $\bm{m} \in \mathbb{R}^{N \times D}$ that adheres to the given keypose and joint trajectory conditions.
In the following, we introduce our novel technical design to effectively inject these conditions.

\subsection{Trajectory ControlNet}
\label{subsec:trajectory_control}

Inspired by the recent success of ControlNet in motion control~\cite{xie2024omnicontrol,zhong2025smoodi,diller2024cg}, we design a trajectory ControlNet $\mathcal{F}_\text{tr}$ to condition the model on joint trajectories.
Specifically, it consists of a trainable copy of the Transformer encoder from the latent diffusion model and each layer is appended with a zero-initialized linear layer $\mathcal{Z}$. 
An independent trajectory encoder $\mathcal{E}_{tr}^{3D}$ is employed to extract the trajectory embedding from the joint trajectory $\mathbf{T}^r_{3D}$.
Considering that controlling joint trajectories requires refining both global spatial and temporal information, we incorporate the joint trajectory embedding into the noised latent $\vz_t$ to create the conditioned noised latent $\vz_{t}'$, which is subsequently fed into the trajectory ControlNet to predict the residual feature $\mathbf{r}$. 

Formally, the whole process can be written as:
\begin{equation}
\vz_{t}' = \vz_t + \mathcal{E}_{tr}^{3D}(\mathbf{T}^r_{3D}), \quad
\mathbf{r} = \mathcal{F}_\text{tr}(\vz_{t}', t, \mathbf{a}).    
\end{equation}
As training progresses, the residual features $\mathbf{r}$ are applied to the corresponding layers in the motion diffusion model $\epsilon_\theta$, implicitly controlling the generated motion.

\subsection{Trajetory-aware Keypose Adapter}
\label{subsec:adaptor}
Unlike the motion in-betweening task ~\cite{cohan2024flexible,harvey2020robust,agrawal2024skel,starke2023motion}, where a globally placed start and end keyposes are known, our single keypose in each sketch frame is drawn independently without the global arrangement.
In essence, the keypose of our task is a representative pose during the action’s execution, providing a more visual and concrete interpretation of the action word.
Moreover, the keypose does not provide global positional information. Instead, it relies solely on its relative timing in relation to the joint trajectories.

In order to inject the keypose information, a straightforward approach is to introduce a keypose ControlNet dedicated to learning keypose constraints. This keypose ControlNet is trained jointly with the trajectory ControlNet, and their residual features are combined before being incorporated into the motion diffusion model.
However, the two controls focus on distinct aspects: the trajectory captures dynamic features and global temporal relationships, while the keypose emphasizes static features and local pose constraints. This divergence creates different levels of learning complexity for each condition, making it difficult to balance their loss weights effectively during training (see Sec.~\ref{subsec:abl_study}).
Moreover, fusing the residual features from multiple ControlNets remains an open challenge, not only in the motion domain~\cite{bian2024motioncraft} but also in the image generation domain~\cite{zhao2024uni,mou2024t2i,qin2023unicontrol}.

To address these challenges, we propose a trajectory-aware keypose adapter, $\mathcal{F}_\text{k}$, that works alongside the trajectory ControlNet $\mathcal{F}_\text{tr}$ to simultaneously inject both constraints.
The keypose adapter accepts the residual output features from the trajectory ControlNet, embedding the keypose condition into the trajectory residual features to achieve deeper interaction and fusion between the two conditions.
Specifically, the keypose adapter $\mathcal{F}_\text{k}$ is also a trainable copy of the Transformer encoder of the latent diffusion model. 
We exploit a keypose encoder $\mathcal{E}_{k}^{3D}$ to extract embeddings from the 3D keypose $\mathbf{K}_{3D}$.
Particularly, 
the keypose complements the action description by providing a detailed interpretation of how the action is executed, we therefore integrate the keypose embedding with the action embedding $\mathbf{a}$ to form a grounded action embedding $\mathbf{a}'$:
\begin{equation}
\label{eq:keypose_embe}
\va' = \va + \mathcal{E}_{k}^{3D}(\mathbf{K}_{3D}).
\end{equation}
Since the trajectory control has already established a global dynamic motion pattern, the keypose adapter only needs to add local static constraints on top of it.
Therefore, the adapter takes the grounded action embedding $\mathbf{a}'$, the noised latent $\vz_t$, and the residual features $\mathbf{r}$ from the trajectory ControlNet as inputs. It learns new residual feature corrections $\mathbf{r}'$ for corresponding layers in \rev{the} motion diffusion model, ensuring the generated motions adhere to both keypose and trajectory constraints.
The process can be expressed as:
\begin{align}
\label{eq:keypose_embe2}
\mathbf{r}' &= \mathcal{F}_\text{k}(\vz_t, t, \mathbf{r}, \va'), \\
\epsilon_{t} &= \epsilon_\theta (\vz_t, t, \mathbf{a}) + \mathcal{Z}(\mathbf{r}').
\end{align}
Through iterative denoising, we obtain the latent code $\vz_0$, which is decoded into a 3D motion using the motion decoder $\mathcal{D}$.

\subsection{Training}
Following~\cite{zhang2023adding}, we freeze the parameters of the motion diffusion model and solely train the trajectory ControlNet and keypose adapter using the standard noise reconstruction loss:
\begin{equation}
\mathcal{L}_\text{recon} = \mathbb{E}_{\epsilon, \vz} \left[ \left\| \epsilon_{\theta}(\vz_t, t, \mathbf{a}, \mathbf{T}^r_{3D},\mathbf{K}_{3D}) - \epsilon \right\|^2_2 \right],
\end{equation}
where $\epsilon \sim \mathcal{N}(0, \mathbf{I})$ represents the ground-truth noise added to $\vz_0$. 
However, solely relying on reconstruction supervision in the latent space is insufficient for the model to effectively learn accurate conditioning on the joint trajectory and keypose.
Therefore, we calculate the $L_{2}$ distance between the joint trajectories of the generated clean motion $\hat{\mathbf{x}}_{0} \in \mathbb{R}^{N \times D}$ at the denoising step $t$ and the ground-truth motion $\mathbf{x}_{0} \in \mathbb{R}^{N \times D}$:
\begin{equation}
\mathcal{L}_\text{tr} = \mathbb{E}_{\hat{\mathbf{x}_0}} \left[ 
\frac{\sum_i \sum_j m_{ij} \| R(\hat{\mathbf{x}}_0)_{ij} - R(\mathbf{x}_0)_{ij} \|_2^2}
{\sum_i \sum_j m_{ij}}
\right],
\end{equation}
where $R(\cdot)$ converts the redundant motion representation into its joint global absolute locations in the space of $\mathbb{R}^{N \times J \times 3}$,
and $m_{ij} \in \{0, 1\}$ is the binary joint trajectory mask at frame $i$ for the joint $j$. 
The generated motion $\hat{\vx}_0$ is obtained by first converting the denoising output latent $\vz_t$ into the predicted clean latent as shown below:
\begin{equation}
\label{eq:oneStep}
\hat{\vz}_0 =  \frac{\vz_t - \sqrt{1 - \alpha_t} \varepsilon_\theta(\vz_t, t, \mathbf{a}, \mathbf{T}^r_{3D},\mathbf{K}_{3D})}{\sqrt{\alpha_t}},
\end{equation}
where $\alpha_t$ denotes the pre-defined noise scale in the forward process of the diffusion model. The predicted clean latent $\hat{\vz}_0$ is then fed into the motion decoder $\mathcal{D}$ to obtain the generated motion.
Similarly, we apply the keypose constraint in the space of $\mathbb{R}^{N \times J \times 3}$ to ensure that the generated motion adheres to the keypose constraints:
\begin{equation}
\mathcal{L}_\text{key} = \mathbb{E}_{\hat{\mathbf{x}_0}} \left[ 
\frac{\sum_i m_{i}' \| R'(\hat{\mathbf{x}}_0)_{i} - R'(\mathbf{x}_0)_{i} \|_2^2}
{\sum_i m_{i}'}
\right],
\end{equation}
where $m_{i} \in \{0, 1\}$ is the binary keypose mask at frame $i$.
To enforce the matching of the ground truth and generated keyposes without considering their global placement, in $R'(\cdot)$, we therefore first use $R(\cdot)$ to convert the motion to the global absolute locations ($\mathbb{R}^{N \times J \times 3}$), then subtract the root position at frame $i$ from each joint's position. 

Overall, the training loss function for the trajectory ControlNet and keypose adapter is defined as follows:
\begin{equation}
\mathcal{L}_\text{gen} = \mathcal{L}_\text{recon} + \lambda_\text{tr} \mathcal{L}_\text{tr} + \lambda_\text{key} \mathcal{L}_\text{key}.
\end{equation}

\section{2D-3D Trajectory and Keypose Alignment}
\label{sec:mapper}

Having the 3D conditional motion generator, instead of formally lifting the 2D keypose and trajectory into the 3D space, 
we propose aligning them in their respective embedding spaces.
Specifically, we train a separate 2D trajectory encoder $\mathcal{E}_{tr}^{2D}$ and keypose encoder $\mathcal{E}_{k}^{2D}$ to align with their frozen 3D counterparts, $\mathcal{E}_{tr}^{3D}$ and $\mathcal{E}_{k}^{3D}$ (Sec. \ref{sec:generator}) after training the motion diffusion model.
Leveraging a shared embedding space between 2D and 3D conditions, we can seamlessly extract 2D embeddings from the storyboard for the conditional motion diffusion model to generate motions.

\mypar{Training.}
Given a batch of paired 2D and 3D embeddings for keyposes and trajectories, denoted as $\mathcal{B}_k = {(\mathbf{s}_{i,k}^{3D}, \mathbf{s}_{i,k}^{2D})}_{i=1}^B$ for keyposes and $\mathcal{B}_{tr} = {(\mathbf{s}_{i,tr}^{3D}, \mathbf{s}_{i,tr}^{2D})}_{i=1}^B$ for trajectories ($B$ is the batch size), we enforce the paired 2D and 3D embeddings to be as close as possible:
\begin{equation}
\mathcal{L}_{\text{match}} = -\frac{1}{B} \sum_{i=1}^B \sum_{y \in \{tr, k\}}\| \mathbf{s}_{i,y}^{3D} - \mathbf{s}_{i,y}^{2D} \|_2^2.
\end{equation}

While the above loss encourages the embeddings of 2D and 3D pairs to be close, it does not guarantee the motions generated based on 2D conditions are realistic.
Therefore, we introduce a noise reconstruction loss for 2D conditions as a motion regularizer:
\begin{equation}
\mathcal{L}_\text{recon} = \mathbb{E}_{\epsilon, \vz} \left[ \left\| \epsilon_{\theta}(\vz_t, t, \mathbf{a}, \mathbf{T}^r_{2D},\mathbf{K}_{2D}) - \epsilon \right\|^2_2 \right].
\end{equation}
\rev{This term couples alignment with generation, yielding 2D embeddings that both align with their 3D embeddings and are seamlessly utilized by the diffusion model for plausible motion generation.}

Additionally, inspired by CLIP-style contrastive learning~\cite{radford2021learning}, we exploit a complementary contrastive learning loss to improve the training efficiency.
The training objective is to maximize the similarity between the embeddings $(\mathbf{s}_{i,k}^{3D}, \mathbf{s}_{i,k}^{2D})$ for keyposes and $(\mathbf{s}_{tr}^{3D}, \mathbf{s}_{tr}^{2D})$ for trajectories, while minimizing the similarity of incorrect pairs.
The learning object can be expressed by:
\begin{equation}
\mathcal{L}_{\text{contrast}} = -\frac{1}{B} \sum_{i=1}^B \sum_{y \in \{tr, k\}} 
\log \frac{\exp(\text{sim}(\mathbf{s}_{i,y}^{3D}, \mathbf{s}_{i,y}^{2D}) / \tau_1)}{\sum_{j=1}^B \exp(\text{sim}(\mathbf{s}_{i,y}^{3D}, \mathbf{s}_{j,y}^{2D}) / \tau_1)},
\end{equation}
where $\text{sim}(\cdot, \cdot)$ is the cosine similarity function, and $\tau_1$ is a temperature hyperparameter. 

The overall training loss function for aligning the embeddings is defined as follows:
\begin{equation}
\mathcal{L}_\text{align} = \mathcal{L}_\text{match} + \lambda_{r} \mathcal{L}_\text{recon} + \lambda_{c} \mathcal{L}_\text{contrast}.
\end{equation}

\section{Inference Motion Generation}
Having the aligned embedding spaces, during inference, we directly input the obtained 2D keypose and joint trajectories (Sec. \ref{sec:overview}) to the multi-conditional motion diffusion model. Furthermore, we apply classifier-free guidance~\cite{ho2022classifier}, and the predicted noise $\epsilon_{\theta}(\vz_t, t, \mathbf{a}, \mathbf{T}^r_{2D}, \mathbf{K}_{2D})$ is computed as follows:
\begin{align}
\label{eql:cfg}
\mathbf{\epsilon}_\theta(\vz_t, t, \mathbf{a}) + 
w_c \big( \epsilon_{\theta}(\vz_t, t, \mathbf{a}, \mathbf{T}^r_{2D}, \mathbf{K}_{2D}) 
- \mathbf{\epsilon}_\theta(\vz_t, t, \mathbf{a}) \big),
\end{align}
where $w_c$ is the scale factor used to control the strength of the conditional guidance.

Additionally, inspired by classifier guidance~\cite{dhariwal2021diffusion} and loss-guided diffusion~\cite{song2023loss,wang2023intercontrol}, we employ inference guidance to enhance the accuracy of following 2D joint trajectories. 
\rev{The idea is to minimize the discrepancy between the projected trajectories of the generated motion and the 2D trajectories. 
See supplementary for the pseudo-codes and inference guidance details.
}

\paragraph{Motion Blending}
\label{sec:blending}
\rev{
After generating a sequence of motion clips from storyboard frames, we apply a post-processing step to create seamless transitions between adjacent clips.
Briefly, given two adjacent motion segments, we first linearly blend them to initialize a transition motion, and then refine it using a deterministic DDIM inversion~\cite{song2020denoising} followed by a guided denoising process. 
We iteratively run the above blending method on all the motion clips within a storyboard composing them into a complete animation.
The implementation details are provided in the supplementary.}

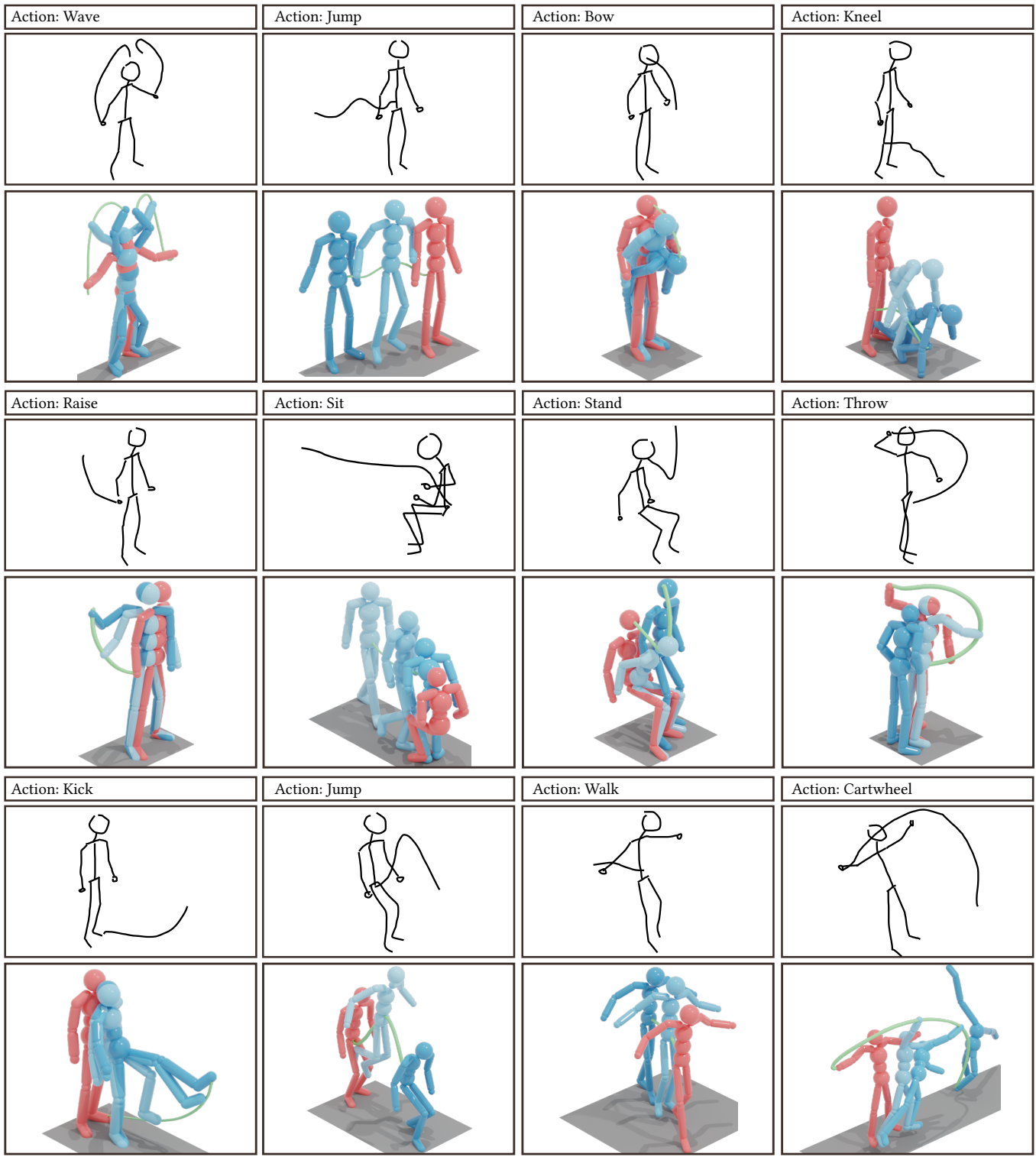
\begin{figure*}[!t]
    \centering
    \begin{overpic}[width=\textwidth]{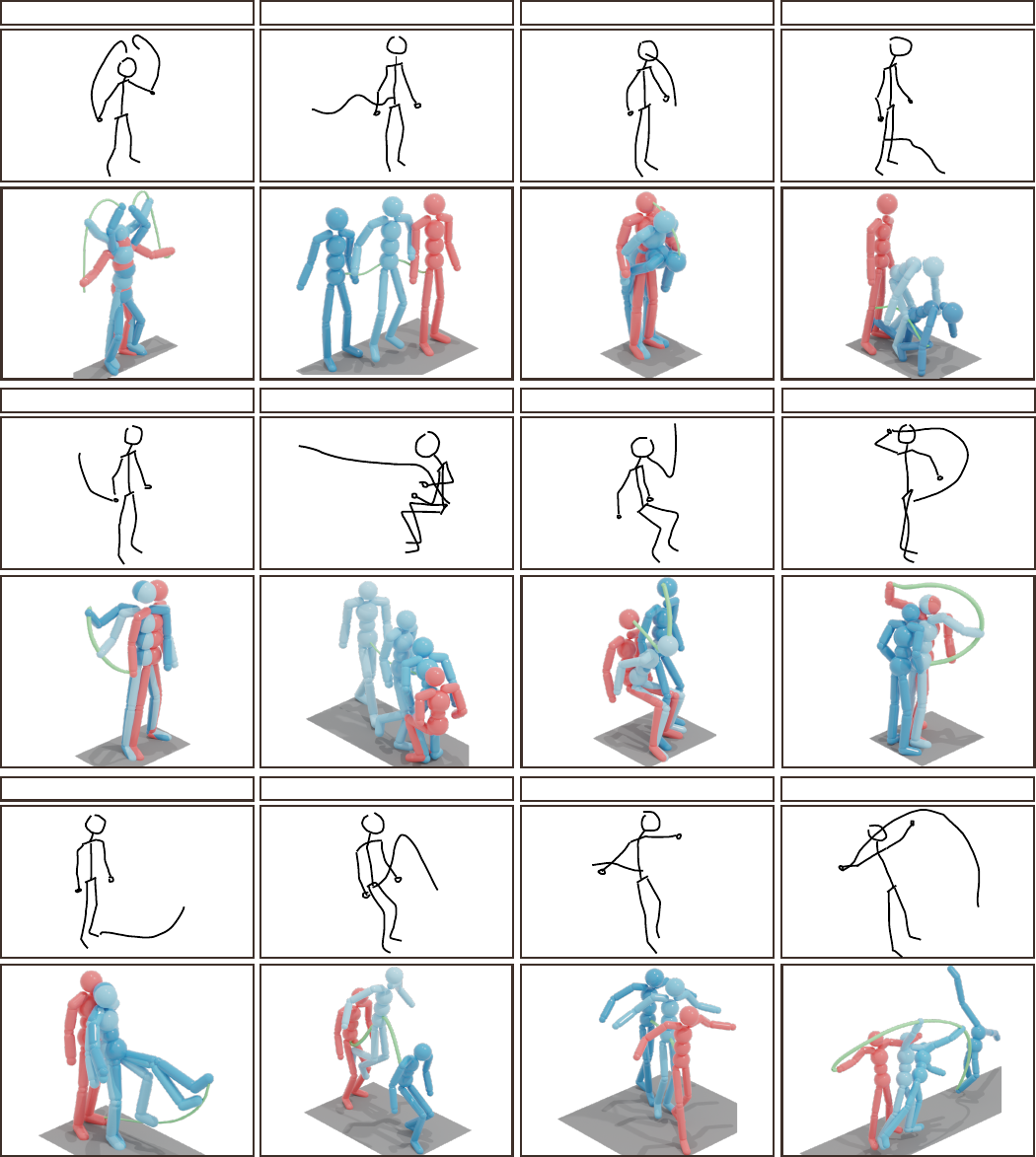} %
        \put(0.6,98.5) {\small Action: Wave}
        \put(23.5,98.5) {\small Action: Jump}
        \put(45.9,98.5) {\small Action: Bow}
        \put(68.4,98.5) {\small Action: Kneel}
        
        \put(0.6,64.9) {\small Action: Raise}
        \put(23.5,64.9) {\small Action: Sit}
        \put(45.9,64.9) {\small Action: Stand}
        \put(68.4,64.9) {\small Action: Throw}
        
        \put(0.6,31.4) {\small Action: Kick}
        \put(23.5,31.4) {\small Action: Jump}
        \put(45.9,31.4) {\small Action: Walk}
        \put(68.4,31.4) {\small Action: Cartwheel}
    \end{overpic}
    \vspace{-2mm}
    \caption{\textbf{Result Gallery}. Three more storyboards are successfully transferred into their high-quality animations. Only the motion clips before blending are shown for clear visualization of keyposes, trajectories, and motion. The complete animation after blending can be seen in the supplemental video. For each clip, we highlight the keypose in red, while the trajectories are spatial curves shown in green.  The motions of ``Kneel'', ``Cartwheel'', and ``Sit'' are usually hard to generate, and our results are of high quality conforming with the sketches. 
    }
    \vspace{-1mm}
    \label{fig:gallary}
\end{figure*}

\section{Results and Discussion}
\label{sec:results}

With our new \name system, we have sketched several storyboards with varying motion complexity. Examples are shown in Figs. \ref{fig:teaser} and \ref{fig:gallary}, with rough sketches (\eg irregular strokes and disproportional body parts), demonstrating vivid motions from `Bow' to the sophisticated `Backflip' and `Kneel'. 
\rev{It is worth noting that the third sketch frame in Fig. \ref{fig:teaser} and the first sketch in Fig. \ref{fig:gallary} include two trajectories for two different joints.}
For better visualization, we only show the motion clips before blending in Fig.~\ref{fig:gallary}. 
Please refer to the \emph{supplemental video} for complete animations.
Extensive comparisons (Sec. \ref{subsec:comp}) and ablation study (Sec. \ref{subsec:abl_study}) validate the effectiveness and core technical design choices of our approach. A user \rev{perceptual} study (Sec. \ref{subsec:user_eval}) also confirms our superior performance. 
We highly encourage the reader to check the supplemental video, which better demonstrates our approach.

\begin{figure}[!tb]
    \centering
    \begin{overpic}[width=0.95\linewidth]{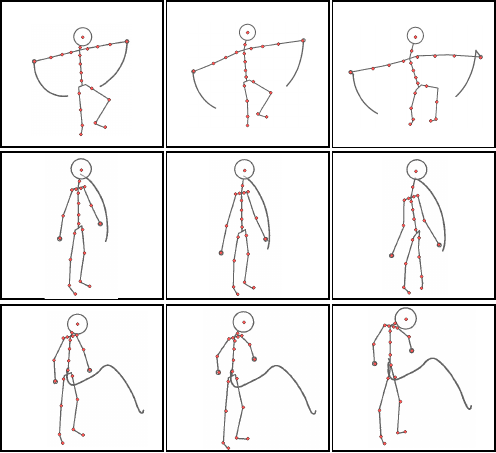} %
        \put(15,-3.5) {\small (a)}
        \put(48,-3.5) {\small (b)}
        \put(81.5,-3.5) {\small (c)}
        \put(2,87) {\small Squat}
        \put(2,57) {\small Bow}
        \put(2,26) {\small Jump}
    \end{overpic}
    \caption{\textbf{\rev{Synthetic} data and its augmentation.} Three examples from our dataset are shown in the three rows. As introduced in Sec.~\ref{sec:overview}, our input is the 2D joint points (\ie red dots), and we draw the stick figure for better visualization. Column (a) demonstrates input data with perfect joint points and body part proportions. While in column (b), random perturbations are applied to the body part proportions (\eg the enlarged neck in ``Bow'' and arms in ``Squat''), resulting in disproportional \rev{keyposes}. A further random perturbation is added to a few joint points in (c) to increase the deviation of joints to mimic the detected joints from user raw sketches.
    }
    \vspace{-3mm}
    \label{fig:data_aug}
\end{figure}

\paragraph{Dataset.}
We train and evaluate our system on the HumanML3D dataset~\cite{Guo_2022_CVPR}, containing 14,646 motions with 44,970 corresponding motion annotations.
Following the processing approach outlined in \rev{~\citet{Guo_2022_CVPR}}, we preprocess the HumanML3D dataset to obtain the redundant motion representations (Sec. \ref{sec:overview}).
Given a motion, we further process it to select a representative keypose corresponding to the detected action word, and project the 3D motion to obtain the 2D keypose and trajectory with data augmentation (\ie camera view augmentation, joint perturbation, and body proportion perturbation). Examples of our synthetic dataset and its augmented variations are shown in Fig.~\ref{fig:data_aug}, and more details of the dataset processing can be found in the supplementary.

\begin{table*}[t]
\centering
\caption{Quantitative analysis of \name (Ours) and three baseline models defined in Sec. \ref{subsec:comp} on the HumanML3D dataset. Evaluation metrics on motion realism, control accuracy, and text-motion match are presented. Following OmniControl~\cite{xie2024omnicontrol}, we report both the average error of all joints (Average) and their random combination (Cross). The best results are highlighted. 
}
\label{tab:compare_with_baseline}
\vspace{-2mm}

\resizebox{0.98\textwidth}{!}{
\begin{tabular}{c|c|cc|cccc|cc}
\toprule[0.25ex]
\multirow{2}{*}{Condition} & \multirow{2}{*}{Method} & \multicolumn{2}{c|}{Realism} & \multicolumn{4}{c|}{Control Accuracy} & \multicolumn{2}{c}{Text-Motion Matching} \\
\cmidrule(lr){3-4} \cmidrule(lr){5-8} \cmidrule(lr){9-10}
 &  & FID $\downarrow$ & Foot Skating $\downarrow$ & MPJPE-2D $\downarrow$ & MPJPE-3D $\downarrow$ & Avg. Err.-2D $\downarrow$ & Avg. Err.-3D $\downarrow$ & MM Dist $\downarrow$ & R-precision (Top-3) $\uparrow$ \\
\midrule
\multirow{4}{*}{Average} 
 & Motion Retrieval & 0.690 & \textbf{0.0640} & 0.0570 & 0.0760 & 0.290 & 0.410 & 4.060 & 0.640 \\
 & Lift-and-Control & 0.979 & 0.0885 & 0.0537 & 0.0712 & 0.261 & 0.340 & 3.297 & 0.752 \\
 & Direct 2D-to-Motion & 2.553 & 0.112 & 0.0403 & 0.0552 & 0.193 & 0.275 & 3.723 & 0.687 \\
 & Ours &  \textbf{0.525} & 0.103 & \textbf{0.0360} & \textbf{0.0478} & \textbf{0.0867} & \textbf{0.134} & \textbf{3.077} & \textbf{0.802} \\
\midrule
\multirow{4}{*}{Cross} 
 & Motion Retrieval & \textbf{0.103} & \textbf{0.0668} & 0.0549 & 0.0730 & 0.3071 & 0.423 & 3.405 & 0.724 \\
 & Lift-and-Control & 0.738 & 0.101 & 0.0505 & 0.0671 & 0.209 & 0.283 & 3.135 & 0.778 \\
 & Direct 2D-to-Motion & 2.310 & 0.123 & 0.0403 & 0.0555 & 0.189 & 0.266 & 3.606 & 0.709 \\
 & Ours &  0.577 & 0.102 & \textbf{0.0329} & \textbf{0.0462} & \textbf{0.0792} & \textbf{0.132} & \textbf{3.042} & \textbf{0.796} \\
\bottomrule[0.25ex]
\end{tabular}
}
\end{table*}

\begin{figure*}[!tb]
    \centering
    \begin{overpic}[width=\textwidth]{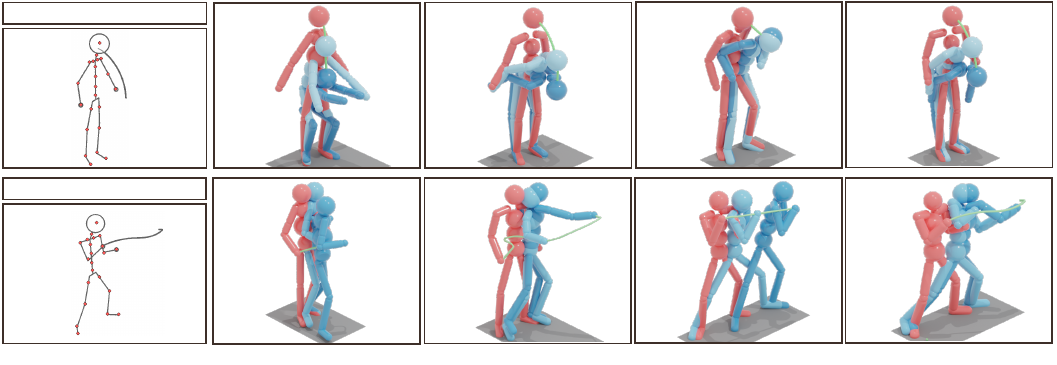} %
        \put(1,32.8) {\small Action: Bow}
        \put(1,16.2) {\small Action: Punch}
        \put(8.2,0) {\small Input}
        \put(25,0) {\small Motion Retrieval}
        \put(45,0) {\small Lift-and-Control}
        \put(63.8,0) {\small Direct 2D-to-Motion}
        \put(89,0) {\small Ours}
    \end{overpic}
    \vspace{-6mm}
    \caption{\textbf{Visual comparison.} Given two frames of the storyboard (\ie ``Kick'' and ``Punch''), corresponding results from competitors and ours are shown. Our results \rev{faithfully adhere to} each frame, while others either have inaccurate joint trajectories or deviated keyposes, leading to unexpected motions. 
    }
    \vspace{-2mm}
    \label{fig:comp}
\end{figure*}

\paragraph{Evaluation metrics.}
We quantitatively evaluate the performance of our framework from three aspects - \emph{motion realism}, \emph{control accuracy}, and \emph{text-motion matching}.
To assess motion realism, we use the Frechet inception distance (FID) to compare the feature distributions of generated motions with real motions.
Furthermore, given the prevalent foot skating issues in kinematics-based motion generation methods, we incorporate the foot skating ratio \cite{karunratanakul2023gmd} as an additional metric to enhance the evaluation of motion quality.
To evaluate the control accuracy, we measure the alignment of both the 2D and 3D conditional keyposes and trajectories, respectively. Note that the resulting 2D keypose and trajectory are obtained by projecting the generated motion.
For the keypose, we measure the mean per joint position error, denoted as MPJPE-2D and MPJPE-3D.
For the trajectory, we similarly measure the average error of the keypose joints, denoted as Avg. Err.-2D and Avg. Err.-3D, respectively.
Although we do not input the 3D conditions at testing time, we report 3D metrics as a complementary measurement.
For text-motion matching, we use motion-retrieval precision (R-precision) to measure the relevancy of the generated motion to its action word. Additionally, we compute multi-modal distance (MM Dist), defined as the average Euclidean distance between the generated motion feature and the corresponding text features.

\paragraph{Implementation Details.}
All models are implemented in Pytorch and trained and tested on a single NVIDIA RTX4090 GPU. Training the trajectory ControlNet and keypose adaptor requires approximately 1000 epochs (12 hours), while aligning the 2D and 3D keypose and trajectory embedding spaces takes around 100 epochs (3 hours). AdamW~\cite{loshchilov2017decoupled} optimizer is exploited for both \rev{trainings} with the learning rate of $1e^{-5}$. 
All hyperparameters in the training are set to $1$.
The weight of classifier-free guidance, $w_c$, is set to $7.5$.
We update $\vz_t$ using inference guidance for $4$ iterations at each denoising step.
At inference time, on average, the algorithm takes around 0.5$s$ to produce a motion clip with 40 frames. The blending of two motion clips takes around 1$s$. Detailed running time breakdowns can be found in the supplementary.

\subsection{Comparison}
\label{subsec:comp}
Instead of evaluating the full algorithm consisting of motion generation and blending, we only consider the core problem of motion clip generation from a single storyboard frame in our comparison. Formally, given the 2D conditional keypose, trajectories, and action word, we compare the generated motion with competitors.

\paragraph{Competitive methods.}
Since no existing method directly addresses our conditional motion generation task, we developed three baseline approaches for comparison.
\begin{itemize}
    \item Motion retrieval:
    since sketch-based motion retrieval works \cite{thorne2004motion,yoo2014sketching} do not have available code, we thus adapt a more recent state-of-the-art learning-based motion retrieval method TMR~\cite{petrovich2023tmr}, which only takes as input the text description to identify the closest motion clip from the dataset. We further develop it to accept additional 2D keyposes and joint trajectories. The resulting method serves as a simple yet effective benchmark for comparison. 
    
    \item Lift-and-control:  
    we trained an additional network to lift 2D keyposes and trajectories to the 3D space. These 3D representations are then fed into the multi-conditional motion diffusion model for motion synthesis. The lifting network is implemented based on MotionBERT~\cite{zhu2023motionbert}.

    \item Direct 2D-to-motion:
    instead of using 3D inputs as surrogates in the motion generator, a multi-conditional motion diffusion model is trained directly on 2D keyposes, joint trajectories, and action words.
    
\end{itemize}

For more details on the implementation of these baselines, please refer to the supplementary.

\paragraph{Evaluation.}
Table~\ref{tab:compare_with_baseline} compares \name with three baseline methods on motion realism, control accuracy, and text-motion matching metrics on our testing dataset.
Following OmniControl~\cite{xie2024omnicontrol}, we report the metrics in terms of the average error over all joints in the first section (denoted as Average), and also the mean error on cross-combinations of all joints, with one combination randomly sampled per test example in the second section (denoted as Cross).
The Motion Retrieval method ($1^{st}$ and $5^{th}$ rows) nearly achieves the best performance on realism metrics, as they utilize real motion clips retrieved from pre-collected datasets. However, their reliance on the diversity and scope of the pre-collected data limits their performance in both control accuracy and text-motion matching metrics. 
Compared to the Direct 2D-to-Motion method ($3^{rd}$ and $7^{th}$ rows), \name demonstrates substantial improvements in control accuracy, leading to more precise and higher quality motions (\eg $5$ times lower FID scores).
The advancements highlight the significance of using 3D inputs as surrogates to guide motion synthesis.
Similarly, as displayed in the $2^{nd}$ and $6^{th}$ rows, the Lift-and-Control method demonstrates reasonable performance across all metrics but is consistently inferior compared to \name. 
Our improvements underscore the advantages of aligning 2D-3D trajectory and keypose embeddings, which effectively capture both spatial and temporal features from the 2D inputs while mitigating potential errors introduced by the lifting step.

The visual comparison is presented in Fig.~\ref{fig:comp}.
The Motion-Retrieval method can obtain a motion in high realism, but it fails to conform with the sketch pose and trajectory. Note that the punching hand of the retrieved motion is even wrong for the action. 
Both the keypose and trajectory of the resulting motions from the Lift-and-Control method are incorrect, due to the inherent inaccuracy of lifting 2D to 3D.
The Direct 2D-to-Motion method produces the worst results, either producing weird motions (\eg ``Bow'') or failing to match the action word. This is mainly because of the considerable domain gap between 2D sketches and 3D animations.
Our results outperform all competitors significantly, the animation is of high quality in terms of motion realism, control accuracy, and text-motion matching.
\emph{These visual differences are best viewed in our supplemental video}.

\begin{table*}[!t]
\centering
\caption{Ablation studies on multi-conditional motion diffusion model design using the HumanML3D dataset. Sec.~\ref{subsec:abl_study} defines all the variants, and we use the \emph{3D keypose} and \emph{3D trajectory} as conditions in this experiment, thus only reporting the control accuracy in its 3D format. Best results are highlighted.}
\label{tab:ablation}
\vspace{-2mm}
\resizebox{0.9\textwidth}{!}{
\begin{tabular}{c|c|cc|cc|cc}
\toprule[0.25ex]
\multirow{2}{*}{Condition} & \multirow{2}{*}{Method} & \multicolumn{2}{c|}{Realism} & \multicolumn{2}{c|}{Control Accuracy} & \multicolumn{2}{c}{Text-Motion Matching} \\
\cmidrule(lr){3-4} \cmidrule(lr){5-6} \cmidrule(lr){7-8}
 &  & FID $\downarrow$ & Foot Skating Ratio $\downarrow$ & MPJPE-3D $\downarrow$ & Avg. Err.-3D $\downarrow$ & MM Dist $\downarrow$ & R-Precision (Top-3) $\uparrow$ \\
\midrule
\multirow{4}{*}{Average} 
 & Single ControlNet & 0.746 & 0.107 & 0.0566 & 0.257 & 3.077 & 0.787 \\
 & Double ControlNet & 0.608 & \textbf{0.0933} & 0.0428 & 0.246 & 3.046 & 0.790 \\
 & Global Keypose & 0.462 & 0.0958 & 0.0413 & 0.190  & 2.932 & 0.806 \\
 & Full Method & \textbf{0.446} & 0.0953 & \textbf{0.0399} & \textbf{0.162} & \textbf{2.884} & \textbf{0.814} \\
\midrule
\multirow{4}{*}{Cross} 
 & Single ControlNet & 0.612 & 0.107 & 0.0541 & 0.210 & 3.007 & 0.796 \\
 & Double ControlNet & 0.433 & 0.0953 & 0.0418 & 0.199 & 2.950 & 0.803 \\
 & Global Keypose & 0.389 & 0.114 & 0.0402 & 0.172 & 2.903 & 0.793 \\
 & Full Method & \textbf{0.370} & \textbf{0.0959} & \textbf{0.0397} & \textbf{0.155} & \textbf{2.897} & \textbf{0.815} \\
\bottomrule[0.25ex]
\end{tabular}
}
\end{table*}

\begin{figure*}[!tb]
    \centering
    \begin{overpic}[width=\textwidth]{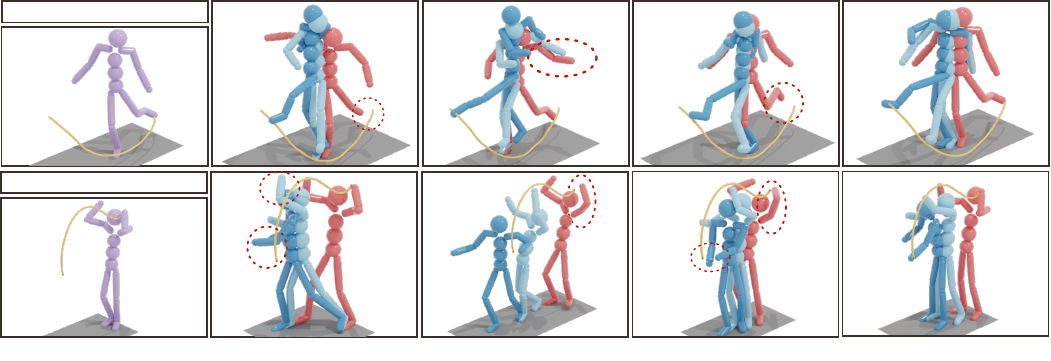} %
        \put(0.8,32.8) {\small Action: Kick}
        \put(0.8,16.5) {\small Action: Throw}
        \put(5,0.5) {\small 3D Conditions}
        \put(24.2,0.3) {\small Single ControlNet}
        \put(44,0.3) {\small Double ControlNet}
        \put(65,0.3) {\small Global Keypose}
        \put(86.1,0.3) {\small Full Method}
    \end{overpic}
    \vspace{-7mm}
    \caption{\textbf{Visual results of the ablation study.} Given the conditional 3D keypose (\ie the purple character) and trajectory (\ie the yellow curve), resulting motions from alternative methods and ours are shown. Note that the input 3D trajectory is overlaid with the produced motions to help better spot the trajectory deviation. The top example demonstrates the poor matching ability of alternative methods regarding the trajectory (see the circled deviations), while the second example mainly displays the inaccurate keyposes from alternative methods. 
    }
    \vspace{-3mm}
    \label{fig:abl_comp}
\end{figure*}
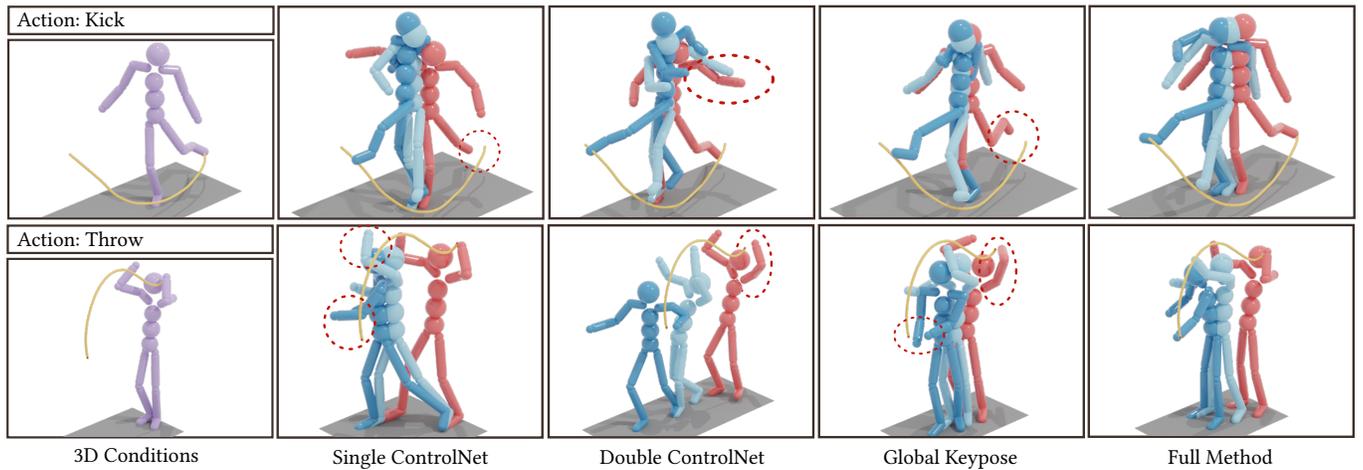

\subsection{Ablation Study}
\label{subsec:abl_study}
We conducted a series of ablation studies to validate the key design choices of our 3D conditional motion generator, with primary focuses on the network modules, leaving the analysis of loss terms in the supplementary. Note that we use the 3D keypose and trajectory in this experiment only to validate the motion generator.

\paragraph{Single ControlNet.}
As stated in Sec. \ref{sec:overview}, the static and local keypose and the dynamic and global trajectory have the same point-based representation. Theoretically, it is possible to assemble them together in a unified matrix, and then feed it to a single ControlNet as the condition for motion generation.

As can be seen in Tab.~\ref{tab:ablation}, this scheme achieves the worst performance across almost all realism and control accuracy metrics.
This significant performance degradation (\eg more than $50\%$ drop off) highlights the limitations of a single ControlNet, which struggles to encode both keypose and trajectory conditions uniformly, especially because keyposes are significantly sparser than trajectories.

\paragraph{Double ControlNet.}
Rather than using an adapter, a separate ControlNet offers a more straightforward approach for incorporating the keypose condition. In this configuration, the keypose feature is still combined with the action word, but the keypose ControlNet operates independently, without receiving residual features from the trajectory ControlNet, keeping the two modules parallel.
The residual features generated by the two ControlNets are added together and subsequently fed into the motion diffusion model.

As shown in Tab.~\ref{tab:ablation}, the double ControlNet scheme performs better than a single ControlNet but worse than ours.
We observe that the two ControlNets compete with each other, resulting in oscillatory training curves. 
In contrast, the adapter focuses on refining the keypose at a specific timestep, leveraging the globally coherent motion provided by the trajectory.
On the other hand, the convergence speeds of the two conditions differ significantly, with keyposes converging nearly ten times faster than joint trajectories.
This disparity poses a significant challenge in dynamically balancing the loss weights for keypose and joint trajectory during training.

\paragraph{Global keypose control.}
In Eq.~\ref{eq:keypose_embe}, we add the keypose embedding to the action word embedding, because we consider keypose as a detailed interpretation of how the action is executed.
To validate the effectiveness of this design, we evaluate the performance of a variant solution where the keypose embedding is added to the noise latent, enabling more global control. Formally, we replace the original Eqs. \ref{eq:keypose_embe} and \ref{eq:keypose_embe2} with $\vz_{t}' = \vz_t + \mathcal{E}_{k}^{3D}(\mathbf{K}_{3D})$ and $\mathbf{r}' = \mathcal{F}_{k}(\vz_{t}', t, \mathbf{r}, \va)$,
while keeping the remaining processes the same.

As shown in Tab.~\ref{tab:ablation}, the full method outperforms the global keypose control scheme on almost all metrics.
Notably, the full method achieves $10\%$ and $4.9\%$ improvements in FID under the Average and the Cross evaluation modes, highlighting the effectiveness of keypose instantiation for interpreting action words.
Furthermore, in terms of trajectory control accuracy (Avg. Err.-3D), the global keypose control scheme shows degradations of $17.2\%$ and $11.0\%$ under the Average and Cross modes, respectively, indicating interference from the keypose on trajectory control. 

A separate visual comparison is presented in Fig.~\ref{fig:abl_comp}, where we demonstrate the resulting motions from all method variants, given the same 3D keyposes and trajectories as the conditions. 
The alternative technical designs fail to explain the sketch keypose or trajectory, producing unsatisfactory motions, which is consistent with the metrics in Tab.~\ref{tab:ablation}.

\begin{figure}[!t]
    \centering
    \begin{overpic}[width=\linewidth]{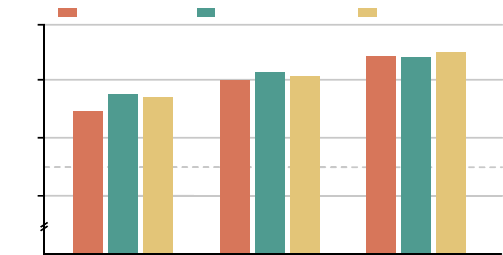} %
        \put(23,-3.5) {\small (a)}
        \put(52,-3.5) {\small (b)}
        \put(81,-3.5) {\small (c)}
        \put(5.5, -0.9) {\small 0}
        \put(4, 10.7) {\small 40}
        \put(4, 22.1) {\small 60}
        \put(4, 33.7) {\small 80}
        \put(2.6,44.7) {\small 100}
        \put(76,47.3) {\footnotesize Keypose accuracy}
        \put(44,47.3) {\footnotesize Trajectory accuracy}
        \put(16.7,47.3) {\footnotesize Motion Realism}
        \put(16,30) {\small 69}
        \put(23,32.8) {\small 76}
        \put(30,32.6) {\small 74}
        \put(45,35.8) {\small 80}
        \put(52.1,37) {\small 84}
        \put(59,36.8) {\small 82}
        \put(74.5,40.6) {\small 88}
        \put(81.2,40.4) {\small 87}
        \put(88.1,41.4) {\small 90}
        \put(0,6.2) {\small \textbf{\rotatebox{90}{\name wins (\%)}}}
    \end{overpic}
    \caption{\textbf{User \rev{perceptual} study results.} The percentage of times our approach is preferred over (a) Motion Retrieval, (b) Lift-and-Control, and (c) Direct 2D-to-Motion is reported. When choosing from a pair of generated motions, users are asked to evaluate three aspects - motion realism, trajectory accuracy, and keypose accuracy. The higher the percentage ($50\%$ is a tie), the better our results.
    }
    \vspace{-3mm}
    \label{fig:user_study}
\end{figure}
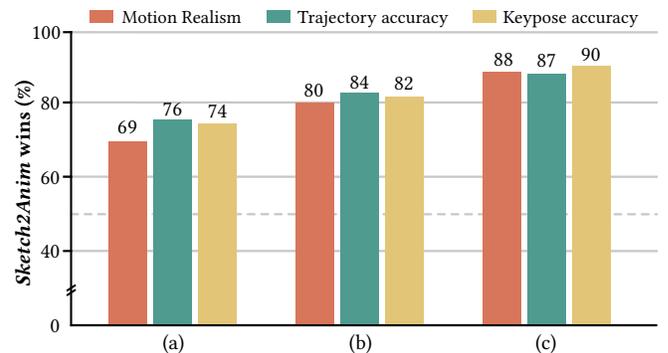

\subsection{User Evaluation}
\label{subsec:user_eval}
We conduct user perceptual studies using pairwise comparisons. 
In each test, participants are presented with a sketch storyboard frame and a pair of motion clips - one produced by \name and the other by a competing method (\ie Motion Retrieval, Lift-and-Control, and Direct 2D-to-Motion).
For each pair of motions, participants are asked to evaluate three aspects - motion realism, trajectory accuracy, and keypose accuracy, as shown in the legend in Fig.~\ref{fig:user_study}.
In total, we have invited $58$ participants, and the percentage of times our approach is preferred over the competing methods is displayed in Fig.~\ref{fig:user_study}.
Our \name performs favorably against Direct 2D-to-Motion for all three aspects, with $88\%$, $87\%$, and $90\%$, respectively.
Similarly, compared with Lift-and-Control, although the percentage our \rev{method wins} drops around $6\%$, it is still considerably high (over $82\%$). 
Lastly, compared to Motion Retrieval, our approach is preferred for motion realism only $69\%$ of the time. While this is above $50\%$ (indicating no tie), the advantage is not substantial. This outcome is understandable, as the retrieved motions from the dataset are captured motions with fewer artifacts, such as foot skating. On the aspects of trajectory and keypose accuracies, our method outperforms Motion Retrieval more than $74\%$ of times. 
Overall, the user perceptual study validates our superior performance and is consistent with observations in the comparison.

\subsection{Application and Discussion}
\label{subsec:discussion}

\begin{figure}[!tb]
    \centering
    \begin{overpic}[width=\linewidth]{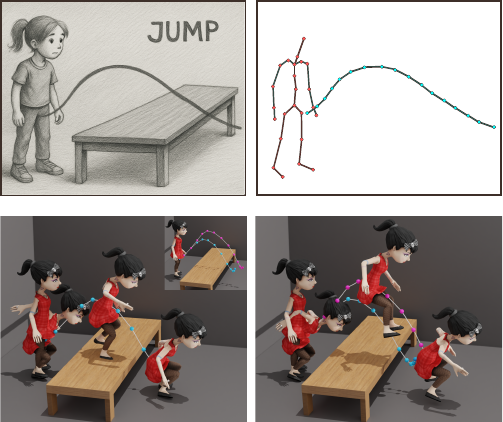} %
        \put(24,42.2) {\small (a)}
        \put(24,-3) {\small (c)}
        \put(72.7,42.2) {\small (b)}
        \put(72.7,-3) {\small (d)}
    \end{overpic}
    \caption{\textbf{3D editing.} Given one frame of a real-world storyboard (a), we preprocess it to obtain the keypose (\ie 2D joint points in red) and the trajectory (\ie curve points in cyan) (b). Our \name produces the animation conditioned on the action word, 2D keypose, and the 2D trajectory (c). Note that we manually model the room with a table for visualization purposes, and the 3D trajectory from the resulting motion is highlighted with cyan points. If the user is unsatisfied with the motion (\eg the foot penetrates the table), the 3D trajectory can be further edited by dragging a few sample points to create a new 3D trajectory (purple points at the top-right corner in (c)), and the updated motion (d) is then re-generated based on the new 3D trajectory condition.
    }
    \vspace{-4mm}
    \label{fig:editing}
\end{figure}
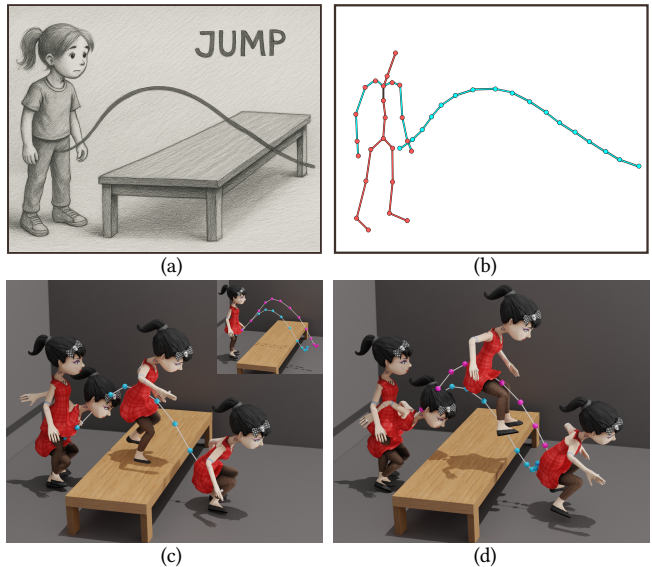

\paragraph{Real-world storyboard and 3D editing}
As stated in Sec.~\ref{sec:overview}, our interface supports simple stick-figure style line segments to ease user burden. However, our algorithm is not limited to only handling simple sketches. As shown in Fig.~\ref{fig:editing} (a), given a real-world sketch storyboard, our pre-processing step can faithfully extract keypose points and trajectory points (Fig.~\ref{fig:editing} (b)). These detected 2D points are conditions consumed by the motion generator to produce the vivid motion (Fig.~\ref{fig:editing} (c)).

As a by-product, our conditional motion generator inherently supports 3D motion editing. As illustrated in Fig.~\ref{fig:editing} (c), after generating the 3D motion from the storyboard frame, the corresponding 3D keyposes and trajectories can be effortlessly extracted. 
If users are unsatisfied with the results, our interface allows point-based keypose and trajectory editing by directly dragging and repositioning the points (see the edited trajectory points in purple), which is very similar \rev{to \citet{agrawal2024skel}}.
Thanks to the technical design of the motion generator, without any modification, our motion generator can take as input the user-edited keypose and trajectories as new conditions to update the motion (Fig.~\ref{fig:editing} (d)).

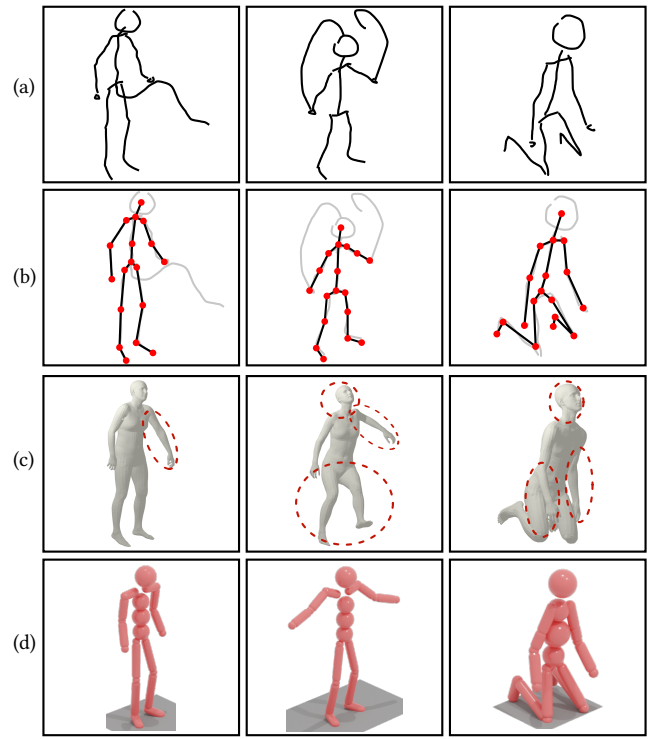
\begin{figure}[!tb]
    \centering
    \begin{overpic}[width=0.94\linewidth]{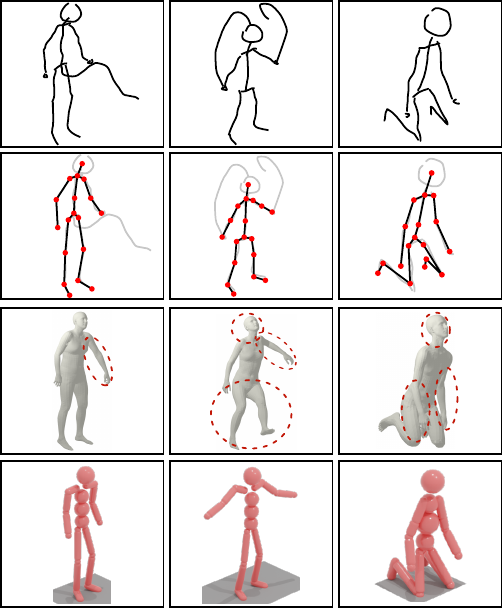} %
        \put(-4, 88) {\small (a)}
        \put(-4, 62) {\small (b)}
        \put(-4,37) {\small (c)}
        \put(-4,12) {\small (d)}
    \end{overpic}
    \vspace{-3mm}
    \caption{\textbf{Joint detection.} Given user rough sketches (a) with crooked and wobbly strokes, and disproportionally placed body parts, we pre-process them using Sketch2Pose to obtain the 2D joints (b). Their 2D joint detection is robust, but their lifted 3D keyposes (c) are problematic. See the highlighted incorrect parts, \eg heads, arms, and legs. Taking as input their 2D joints as a condition, we can produce high-quality motions, whose corresponding 3D keyposes (d) faithfully conform with input sketches.
    }
    \vspace{-4mm}
    \label{fig:joint_detection}
\end{figure}

\paragraph{Robustness of joint detection}
In Sec.~\ref{sec:overview}, the pre-processing with Sketch2Pose is introduced. Our method is highly dependent on the success of joint detection, especially since the user's raw sketches are often irregular, rough, and \rev{have disproportionate} body parts. 
We thus particularly validate the robustness of joint detection from Sketch2Pose in our storyboard sketch scenario.
A few examples are presented in Fig. \ref{fig:joint_detection}. Even though the strokes are crooked and wobbly (Fig. \ref{fig:joint_detection}(a)), Sketch2Pose can successfully and robustly detect the joints (Fig. \ref{fig:joint_detection}(b)), but their 3D keypose estimation (Fig. \ref{fig:joint_detection}(c)) is not \rev{reliable}, which is the main reason we only take as input their 2D keypose instead of the lifted 3D keypose in our algorithm. 
As a comparison, we extract 3D keyposes from our generated motions corresponding to input sketches and display them in Fig. \ref{fig:joint_detection}(d), where our 3D keyposes conform to input sketches faithfully. 

\begin{figure}[!t]
    \centering
    \begin{overpic}[width=0.97\linewidth]{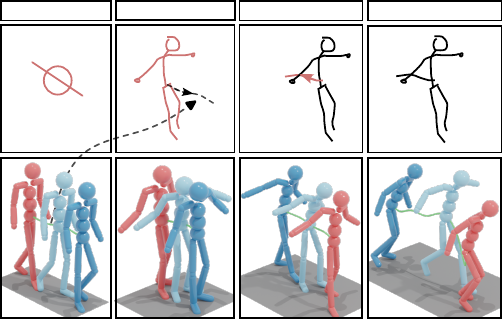} %
        \put(9,-3.5) {\small (a)}
        \put(33,-3.5) {\small (b)}
        \put(60,-3.5) {\small (c)}
        \put(86,-3.5) {\small (d)}
        \put(1,60.2) {\small \textcolor{mypink1}{Action: Walk}}
        \put(24,60.2) {\small Action: Walk}
        \put(49,60.2) {\small Action: Walk}
        \put(74,60.2) {\small \textcolor{mypink1}{Action: Jump}}
        \put(36,48) {\footnotesize \textcolor{gray}{forward}}
        \put(50,44) {\footnotesize \textcolor{gray}{backward}}
    \end{overpic}
    \caption{\textbf{Impact of conditions.} From left to right, we gradually add or change one condition for our method when generating motions at inference time. We highlight the change with the pink color. (a) Only the action word ``Walk'' is input as the condition. (b) The action word and the generated forward root joint trajectory are re-used. A new sketch keypose is provided. (c) The forward trajectory is replaced by a backward trajectory. (d) The action word is replaced by ``Jump''. 
    }
    \label{fig:impact_condition}
\end{figure}

\begin{figure}[!t]
    \centering
    \begin{overpic}[width=0.98\linewidth]{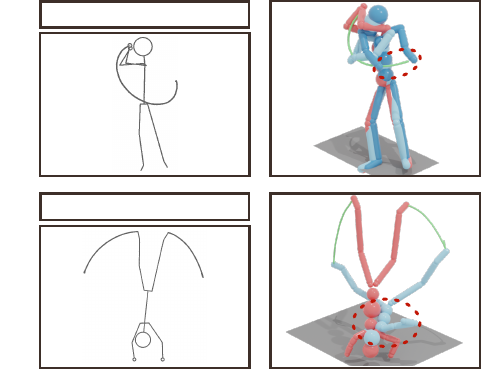} %
        \put(10,70.6) {\small Action: Golf}
        \put(10,31.6) {\small Action: HandStand}
        \put(26, -3) {\small Input}
        \put(66, -3) {\small Output Motion}
        \put(2.5,55) {\small (a)}
        \put(2.5,16) {\small (b)}
    \end{overpic}
    \caption{\textbf{Limitations.} (a) Our method does not consider character-object interaction, thus the two hands do not hold the golf club at the ending keypose. (b) Without the physical constraints, even if the foot trajectories are correct, the body folds and floats in the air at the end of the motion. The errors are highlighted with dashed red circles.
    }
    \vspace{-3mm}
    \label{fig:limitation}
\end{figure}
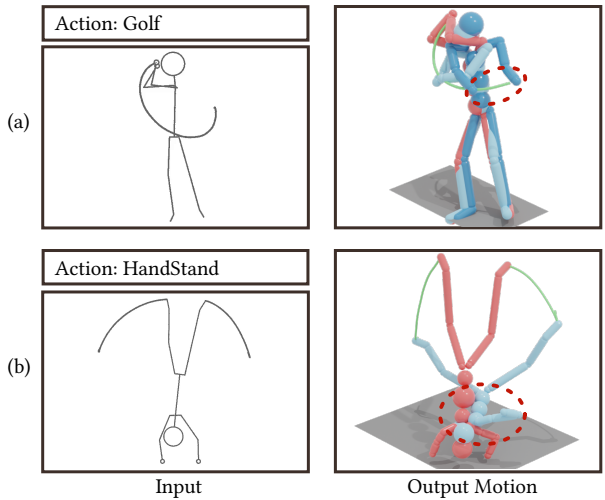

\paragraph{Impact of conditions}
We have ablated our generator design and the training and inference loss terms in Sec. \ref{subsec:abl_study} and the supplementary, respectively. Here, we further validate the impact of our conditions. To this end, we start with an action word to generate a motion and gradually add or change one condition each time when generating new motions at inference time.
The results can be seen in Fig. \ref{fig:impact_condition}, where in subfigure (a), given only the action word ``Walk'' as the condition, our method produces a satisfactory walk motion with two hands down. 
We then re-use the action word as well as the generated root joint trajectory in (a) and take as input a new sketch keypose (Fig. \ref{fig:impact_condition}(b)) to produce an updated motion with the two hands up conforming with the sketch.
Note that the root joint trajectory in Fig. \ref{fig:impact_condition}(a) is forward, see the dashed lines.
We further change the root joint trajectory to a new backward sketch (Fig. \ref{fig:impact_condition}(c)), and the generated motion is successfully adjusted.
Lastly, we keep the sketch keypose and the backward trajectory, but change the action word from ``Walk'' to ``Jump'' (Fig. \ref{fig:impact_condition}(d)). The resulting motion is a mild jumping, constrained by the relatively flat sketch trajectory. 
Because of the conflicts between the trajectory and the action word, deviations of keypose and trajectory from the sketches are reasonably observed, while the motion is still satisfying.   

\paragraph{Limitations}
Our method has a few limitations. 
Firstly, we do not consider character-object interaction in our algorithm design. Thus, in Fig.~\ref{fig:limitation}(a), even \rev{though} the storyboard frame demonstrates the animation of playing golf, the two hands of the ending keypose are separated, without holding the golf club.
Secondly, our algorithm does not incorporate physical constraints, leading to physically incorrect keyposes. For instance, in Fig.~\ref{fig:limitation}(b), the body folds and floats in the air at the end of the ``HandStand'' motion. Extra physical constraints (\eg a loss term) extracted from the action word might solve this problem. 
Thirdly, although we have validated the robustness of joint detection, we could not explore all the variants of user sketches. Clearly, Sketch2Pose~\cite{brodt2022sketch2pose} might fail if the drawing deviates too much from a reasonable human character. This can be addressed by re-drawing some strokes or manually positioning misplaced joint points in the interface.

\section{Conclusion and Future Work}

We have presented \name, the \emph{first} approach for transferring \rev{a} sketch storyboard into its high-quality 3D animation. We solve this problem from the perspective of conditional motion generation, and the key idea is to exploit the informative 3D keypose and trajectory during training of the motion diffusion model to enable precise control, while directly inputting the 2D keypose and trajectory during inference. This is achieved by our dedicated neural mapper to align the 2D keypose and trajectory to their 3D counterparts in the shared embedding spaces. We have extensively evaluated the superior performance of our approach with a comparison, an ablation study, and a user \rev{perceptual} study. We further demonstrated the robustness and flexibility of our method on real-world sketch storyboards and keypose- or trajectory-based motion editing applications.

\paragraph{Future work.} There are a few inspiring directions worth exploring in the future.
\begin{itemize}
    \item[(1)] Speed control: besides keypose and trajectory strokes, over-sketched speed lines are also widely used in storyboard sketches \cite{choi2012retrieval}. For example, a dense speed line pattern conveys a fast motion, providing hints for the animation timeline control. We plan to include a speed line detection and understanding module and inject the extra constraints into the motion generation model. \rev{See more discussion about speed control in the supplementary.}
    
    \item[(2)] Scene recovery: industrial \rev{storyboards} usually \rev{contain} scene strokes representing the surrounding objects and environment (see Fig.~\ref{fig:editing}(a)). 
    It is beneficial if the scene reconstruction and character animation can be considered together with mutual spatial constraints to each other. 
    For instance, if the table in the room is generated together with the motion, it can provide \rev{a} minimum height constraint for the foot when jumping to prevent penetration and collision between the foot and the table.
\end{itemize}

\begin{acks}
The authors would like to thank the reviewers for their valuable suggestions, Shuyuan Zhang for crafting the animation in Fig. \ref{fig:workflow} and the video using the traditional 3D animation workflow in Blender, and Adrien Bousseau, Hakan Bilen for proofreading earlier drafts of the paper. CL was supported by a gift from Adobe. 
YX was supported by the Apple Scholars in AI/ML PhD fellowship.
\end{acks}

\bibliographystyle{ACM-Reference-Format}
\bibliography{main}

\clearpage
\section*{Supplemental Material}

\appendix

\setcounter{table}{0}
\renewcommand{\thetable}{A\arabic{table}}
\setcounter{figure}{0}
\renewcommand{\thefigure}{A\arabic{figure}}

In this supplementary material, we provide additional details about dataset processing, user \rev{evaluation}, the implementation of our approach and all competitors. Furthermore, we include more ablation study results to analyze the impact of each loss term.
We provide a \emph{supplemental video}, which we encourage the reviewers to watch since motion is critical in our results, and this is hard to convey in a static document.

\section{More ablation studies}
In this section, we evaluate the impact of each loss term, when training both the multi-conditional motion diffusion model and the neural mapper to align 2D-3D trajectory and keypose embeddings, respectively.
\emph{Because inference guidance can be seen as a post-optimization process, neither variant employs inference guidance in these ablation studies}.

Firstly, we exclude the spatial keypose and trajectories constraint term when training our multi-conditional motion diffusion model, denoted as w/o $\mathcal{L}_\text{key}\&\mathcal{L}_\text{tr}$.
As shown in Table~\ref{tab:abation_multi_model}, although introducing spatial constraints affects the FID metric, it significantly improves control accuracy, with MPJPE-3D improving by 18.37\% and Avg. Err-3D improving by 
\textbf{46.58\%} in the Average setting, and by 16.07\% and \textbf{42.59\%}, respectively, in the Cross setting.
The trade-off between control constraints and FID metrics is also reported in ~\cite{dai2025motionlcm,zhong2025smoodi}.
For the text-motion matching metric, it degrades a little bit. 

Then, we exclude each loss term in the alignment process.
First, we exclude the \emph{reconstruction} term in the loss function, denoted as w/o $\mathcal{L}_\text{recon}$.
Comparing the results in the \(1^{\text{st}}\) and \(4^{\text{rd}}\) rows in both the Average and Cross settings in Tab.~\ref{tab:ablation_algin}, we observe that removing $\mathcal{L}_\text{recon}$ leads to a considerable degradation in the FID metric, with a \textbf{32.81\%} and \textbf{23.12\%} drop for the Average and Cross settings, respectively. The other two metrics degrade as well. 
When We exclude the \emph{match} term (denoted as w/o $\mathcal{L}_\text{match}$), we observe similar performance change, e.g., the large FID degradation, and the mild drop of the text-motion matching metric.  
Both the \emph{recosntruction} and \emph{match} terms play essential roles in the alignment training.
The \emph{contrast} loss complements the above two terms and further improves the performance to some extent.

\begin{table*}[t]
\centering
\caption{Ablation study of the loss terms of our multi-conditional motion diffusion model using the HumanML3D dataset. Refer to the main paper for the definition of metrics and the Average and Cross evaluation settings.}
\label{tab:abation_multi_model}
\vspace{-3mm}
\renewcommand{\arraystretch}{1.0}
\resizebox{0.8\textwidth}{!}{
    \begin{tabular}{c|c|cc|cc|cc}
    \toprule[0.25ex]
    \multirow{2}{*}{Joint} & \multirow{2}{*}{Method}  & \multicolumn{2}{c|}{Realism} & \multicolumn{2}{c|}{Control Accuracy} & \multicolumn{2}{c}{Text-Motion Matching} \\
    \cmidrule(lr){3-4} \cmidrule(lr){5-6} \cmidrule(lr){7-8}
    &  & FID $\downarrow$ & Foot skating ratio $\downarrow$ & MPJPE-3D $\downarrow$ & Avg. Err.-3D $\downarrow$ & MM Dist $\downarrow$ & R-precision (Top-3) $\uparrow$ \\
    \midrule
    
    \multirow{2}{*}{Average} & w/o $\mathcal{L}_\text{key} \& \mathcal{L}_\text{tr}$ & \textbf{0.291} & \textbf{0.0858} & 0.0490 & 0.307 & \textbf{2.908} & \textbf{0.823} \\
    & Ours & 0.451 & 0.094 & \textbf{0.040} & \textbf{0.164} & 2.934 & 0.810 \\
    \midrule

    \multirow{2}{*}{Cross} & w/o $\mathcal{L}_\text{key} \& \mathcal{L}_\text{tr}$ &  \textbf{0.271} & \textbf{0.0895} & 0.0473 & 0.270 & \textbf{2.858} & \textbf{0.824} \\
    & Ours & 0.370 & 0.0959 & \textbf{0.0397} & \textbf{0.155} & 2.897 & 0.815 \\
    \bottomrule[0.25ex]
    \end{tabular}
}
\end{table*}

\begin{table*}[!ht]
\centering
\caption{Ablation study of the loss terms of our neural mapper to align 2D-3D embeddings using the HumanML3D dataset. Refer to the paper for the definition of the reported metrics and the Average and Cross evaluation settings. Note that the statistics of Ours' are different with the numbers in Table 1 in the main paper, because we did not include inference guidance in this experiment.}
\label{tab:ablation_algin}
\vspace{-3mm}
\renewcommand{\arraystretch}{1.0}
\resizebox{0.98\textwidth}{!}{
    \begin{tabular}{c|c|cc|cccc|cc}
        \toprule[0.25ex]
        \multirow{2}{*}{Joint} & \multirow{2}{*}{Method}  & \multicolumn{2}{c|}{Realism} & \multicolumn{4}{c|}{Control Accuracy} & \multicolumn{2}{c}{Text-Motion Matching} \\
        \cmidrule(lr){3-4} \cmidrule(lr){5-8} \cmidrule(lr){9-10}
        &  & FID $\downarrow$ & Foot Skating $\downarrow$ & MPJPE-2D $\downarrow$ & MPJPE-3D $\downarrow$ & Avg. Err.-2D $\downarrow$ & Avg. Err.-3D $\downarrow$ & MM Dist $\downarrow$ & R-precision (Top-3) $\uparrow$ \\
        \midrule
        
        \multirow{4}{*}{Average} & w/o $\mathcal{L}_\text{recon}$ &  0.768 & 0.120 & 0.0391 & 0.0539 & 0.184 & 0.266 & 3.252 & 0.754 \\
        & w/o $\mathcal{L}_\text{contrast}$  &  0.544 & \textbf{0.0994} & 0.0370 & 0.0509 & 0.162 & 0.232 & 3.208 & 0.743 \\
        & w/o $\mathcal{L}_\text{match}$  &  0.722 & 0.117 & 0.0369 & 0.0509 & 0.172 & 0.247 & 3.232 & 0.757 \\
        & Ours  &  \textbf{0.516} & 0.108 & \textbf{0.0361} & \textbf{0.0498} & \textbf{0.158} & \textbf{0.227} & \textbf{3.128} & \textbf{0.769} \\
        \midrule
        
        \multirow{4}{*}{Cross} & w/o $\mathcal{L}_\text{recon}$  &  0.653 & 0.116 & 0.0384 & 0.0528 & 0.181 & 0.254 & 3.141 & 0.781 \\
        & w/o $\mathcal{L}_\text{contrast}$  &  0.534 & \textbf{0.0998} & 0.0370 & 0.0508 & 0.160 & 0.235 & 2.984 & 0.794 \\
        & w/o $\mathcal{L}_\text{match}$  &  0.663 & 0.115 & 0.0365 & 0.0502 & 0.173 & 0.243 & 3.138 & 0.780 \\

        & Ours   & \textbf{0.502} & 0.107 & \textbf{0.0357} & \textbf{0.0493} & \textbf{0.159} & \textbf{0.225} & \textbf{2.888} & \textbf{0.815} \\
        \bottomrule[0.25ex]
    \end{tabular}
}
\end{table*}

\section{Dataset Processing}
We train and evaluate our system on the HumanML3D dataset~\cite{Guo_2022_CVPR}, containing 14,646 motions with 44,970 corresponding motion annotations.
Following the processing approach outlined in ~\cite{Guo_2022_CVPR}, we preprocess the HumanML3D dateset to obtain the redundant motion representations.
To effectively select a representative keypose from a given motion sequence, we first use NLP tools~\cite{Gardner2017AllenNLP} to extract the \emph{first} verb as the action word from the text description and form a simplified text description with the template of \emph{`a person [action\_word]'}. Next, we slide a window over the motion sequence to compute similarity scores between each motion segment and the simplified sentence using TMR~\cite{petrovich2023tmr}.
The motion segment with the highest similarity score is identified as the candidate group of keyposes, from which individual keyposes are randomly selected during training.
During the evaluation, we exploit the first keypose from the candidate group as the keypose.

To obtain 2D keyposes and joint trajectories resembling user sketches extracted from storyboards, we begin by orthographically projecting the 3D motion to generate the corresponding 2D motion.
From this projected data, we extract the 2D keyposes and joint trajectories. Specifically, we focus on six commonly used end-effector joints - \emph{pelvis, left foot, right foot, head, left wrist}, and \emph{right wrist}, to extract trajectories.
In order to mimic the irregularity of real user sketches at testing time, we employ the following three data augmentation strategies:
\begin{itemize}
    \item {\it Camera view augmentation}: 
    The camera parameter $p$ comprises the scale $s$ and the Euler angles $v = (v^{\text{pitch}}, v^{\text{yaw}}, v^{\text{roll}})$. During training, $v^{\text{roll}}$ is fixed at 0, while $v^{\text{pitch}}$ is randomly sampled from $[0^\circ, 30^\circ]$, $v^{\text{yaw}}$ from $[-45^\circ, 45^\circ]$, and $s$ from $[0.8, 1.2]$. This procedure is crucial for training the network to consistently map similar poses, despite variations in 2D scale and viewpoint, to the same (local) 3D representation.
    \item {\it Random joint perturbation}: Since user sketches often contain rough strokes (\eg crooked lines),
    we add Gaussian noise with a standard deviation of $0.02$ to the 2D joints to better match real-world drawing scenarios.
    \item {\it Body proportion perturbation}: other than the irregularity of a single stroke, the user sketches might have incorrect body proportions, we thus randomly scale selective body parts: leg, spine, and arm, in the 2D projected keypose by a factor $s$, sampled from the range $[0.6, 1.6]$.
\end{itemize}

\section{The details of Inference Guidance}
\rev{
The core of inference guidance is an analytic function.
We optimize this analytic function using the second-order optimizer L-BFGS~\cite{liu1989limited}, which better aligns with the desired trajectory and achieves quicker convergence compared to first-order based methods~\cite{xie2024omnicontrol,pinyoanuntapong2024controlmm}:
\begin{align}
\label{eq:inference_guidance}
\epsilon_{\theta}(\vz_t, t, \mathbf{a}, \mathbf{T}^r_{2D}, \mathbf{K}_{2D}) &= 
\epsilon_{\theta}(\vz_t, t, \mathbf{a}, \mathbf{T}^r_{2D}, \mathbf{K}_{2D}) \notag \\ 
&- \tau_2 \cdot \mathbf{H}^{-1} \nabla_{\vz_t} G(\vz_t, t, \vv, \mathbf{T}^r_{2D}), \\
G(\vz_t, t, \vv, \mathbf{T}^r_{2D}) &= 
\frac{\sum_{i,j} m_{ij} \| P(R(\hat{\mathbf{x}}_0)_{ij}, \vv) - \mathbf{T}^r_{2D} \|_2^2}
{\sum_{i,j} m_{ij}},
\end{align}
where $\tau_2$ controls the strength of the guidance, and $P(\vx, \vv)$ represents the projection of the motion $\vx$ under the camera view $\vv$. The term $\mathbf{H}^{-1}$ denotes the approximate inverse Hessian matrix used in the L-BFGS optimizer~\cite{liu1989limited}.
Because the generated motion aligns closely with the desired 2D keyposes, we do not adopt similar inference guidance for the keypose condition.

\paragraph{Algorithm pseudo-code}
We list the pseudo-code of our algorithm during inference in Algo. \ref{alg:inference}. 
Note that the subscript ${2D}$ and the guidance term show the full algorithm inference (Sec. 6 in the paper), while the subscript ${3D}$ indicates the training/inference process of the 3D conditioned motion generator (Sec. 4 in the paper).

\newcommand\mycommfont[1]{\footnotesize\textcolor{blue}{#1}}
\SetCommentSty{mycommfont}

\begin{algorithm}[!tb]
\caption{{\name}'s inference}
\label{alg:inference}

\textbf{Require:} A motion diffusion model $M$ with parameters $\theta_M$, a Trajectory ControlNet $\mathcal{F}_{tr}$ with parameters $\theta_{tr}$, and a KeyPose Adapter $\mathcal{F}_{k}$ with parameters $\theta_{k}$. The inputs include 2D/3D keyposes $\mathbf{K}_{2D/3D}$, 2D/3D joint trajectories $\mathbf{T}^r_{2D/3D}$, camera view $\vv$, and action word embeddings $\mathbf{a}$, which are obtained from the action word $\mathbf{W}_a$ via CLIP.

$\vz_T \sim \mathcal{N}(\mathbf{0}, \mathbf{I})$ \tcp*[l]{Sample from pure Gaussian distribution}

\For{$t = T$ to $1$}{
    $\{\vr\} \leftarrow \mathcal{F}_{tr}(\vz_t, t, \va, \mathbf{T}^r_{2D/3D}; \theta_{tr})$ \tcp*[l]{Trajectory ControlNet}
    
    $\{\vr'\} \leftarrow \mathcal{F}_{k}(\vz_t, t, \{\vr\}, \va, \mathbf{K}_{2D/3D}; \theta_{k})$ \tcp*[l]{Keypose Adapter}
    
    $\epsilon_t \leftarrow M(\vz_t, t, \va, \{\vr'\}; \theta_M)$ \tcp*[l]{Motion diffusion model}
    
    \tcp*[h]{Inference guidance}\\
    \If{input condition is 2D}{
        \For{$k = 1$ to $K$}{
            $\epsilon_t = \epsilon_t - \tau \cdot \mathbf{H}^{-1} \nabla_{\vz_t} G(\vz_t, t, \mathbf{T}^r_{2D}, \vv)$ \tcp*[l]{2D guidance}
        }
    }
    
    $\vz_{t-1} \sim \mathcal{S}(\vz_t, \epsilon_t, t)$ \tcp*[l]{$\mathcal{S}(\cdot,\cdot,\cdot)$: DDIM sampling~\shortcite{dhariwal2021diffusion}}
}

$\vx_0 = \mathbf{D}(\vz_0)$

\Return $\vx_0$

\end{algorithm}

\begin{table*}[!tb]
\centering
\caption{Ablation study of the inference guidance using the HumanML3D dataset.}
\label{tab:ablation_inference_guidance}
\vspace{-3mm}
\resizebox{0.98\textwidth}{!}{
    \begin{tabular}{c|c|cc|cccc|cc}
        \toprule[0.25ex]
        \multirow{2}{*}{Joint} & \multirow{2}{*}{Method}  & \multicolumn{2}{c|}{Realism} & \multicolumn{4}{c|}{Control Accuracy} & \multicolumn{2}{c}{Text-Motion Matching} \\
        \cmidrule(lr){3-4} \cmidrule(lr){5-8} \cmidrule(lr){9-10}
        &  & FID $\downarrow$ & Foot Skating $\downarrow$ & MPJPE-2D $\downarrow$ & MPJPE-3D $\downarrow$ & Avg. Err.-2D $\downarrow$ & Avg. Err.-3D $\downarrow$ & MM Dist $\downarrow$ & R-precision (Top-3) $\uparrow$ \\
        \midrule

        \multirow{3}{*}{Average} 
        & w/o inf. guidance  &  \textbf{0.516} & 0.108 & 0.0361 & 0.0498 & 0.158 & 0.227 & 3.128 & 0.769 \\
        & w/ $1^{st}$ order grad.  &  0.520 &0.105 & 0.0361 & 0.0490 & 0.102 & 0.183 & 3.093 & 0.797 \\
        & w/ $2^{nd}$ order grad.  &  0.525 & \textbf{0.103} & \textbf{0.0360} & \textbf{0.0478} & \textbf{0.0867} & \textbf{0.134} & \textbf{3.077} & \textbf{0.802} \\
        \midrule

        \multirow{3}{*}{Cross} 
        & w/o inf. guidance  & \textbf{0.502} & \textbf{0.107} & 0.0357 & 0.0493 & 0.159 & 0.225 & \textbf{2.888} & \textbf{0.815} \\
        & w/ $1^{st}$ order grad.  &  0.596 & 0.104 & 0.0350 & 0.0471 & 0.0981 & 0.184 & 2.906 & 0.801 \\
        & w/ $2^{nd}$ order grad.  &  0.577 & 0.102 & \textbf{0.0329} & \textbf{0.0462} & \textbf{0.0792} & \textbf{0.132} & 3.042 & 0.796 \\
        \bottomrule[0.25ex]
    \end{tabular}
}
\end{table*}

\paragraph{Evaluation}
In this work, we employ inference guidance as a post-optimization step to ensure the generated motion better follows the given 2D joint trajectory.
As illustrated in the $1^{st}$ and $3^{rd}$ rows in Tab.~\ref{tab:ablation_inference_guidance}, using the second-order gradient significantly improves the trajectory control accuracy, with Avg. Err.-2D decreasing by $45.13\%$ in the Average setting and $50.19\%$ in the Cross setting, and Avg. Err.-3D decreasing by $40.97\%$ and $41.33\%$, respectively.
Moreover, previous methods~\cite{xie2024omnicontrol, pinyoanuntapong2024controlmm, karunratanakul2023gmd} primarily rely on first-order gradient-based approaches to ensure that the generated motion adheres to the given trajectories.
We thus report the metrics by using first-order gradients when applying inference guidance.
The results in Tab.~\ref{tab:ablation_inference_guidance} demonstrate that second-order methods achieve better performance than first-order methods across all metrics.

}

\section{The details of Motion Blending}
\rev{
In this section, we describe the full details of our inversion-based motion blending and evaluate its performance by a statistical comparison.

\paragraph{Implementation.} 
Given two adjacent motion clips, $\hat{\mathbf{x}}^{p}_{0}$ and $\hat{\mathbf{x}}^{q}_{0}$, we select a segment of length $l$ at the junction of their start and end as the motion transition.
We initialize a transition motion $\mathbf{x}_{0}^{l}$ using linear blending and compose it with adjacent clips to form $\mathbf{x}^{p+q}_{0}$. Subsequently, we apply the deterministic DDIM inversion process \rev{\cite{song2020denoising}} to obtain the noised latent code $\mathbf{z}_T^{\text{Inv}}$ for the concatenated motion.
The reverse process can be represented at step $t$ as:
\begin{equation}
\vz_{t+1} = \sqrt{\frac{\omega_{t+1}}{\omega_t}} \left( \vz_t + \left( \sqrt{\frac{1}{\omega_{t+1}}} - 1 \right) - \left( \sqrt{\frac{1}{\omega_t}} - 1 \right) \right) \cdot \varepsilon_{\theta}(\vz_t; t, \vc,\emptyset),
\end{equation}
where $\omega$ represents the noise scale.
$\vz_T^{Inv}$ can be obtained at the last reverse step $T$. 
During inference, we denoise $\mathbf{z}_T^{Inv}$ using a combination of classifier-free guidance and second-order inference guidance as follows:
\begin{align}
\mathbf{\epsilon}_\theta(\vz_t, t, \mathbf{a}_{p+q}) &= \mathbf{\epsilon}_\theta(\vz_t, t, \mathbf{a}_{p+q}) - 
\tau_3 \cdot \mathbf{H}^{-1} \nabla_{\vz_t} G_m(\vz_t, t, \hat{\mathbf{x}}^{p}_{0}, \hat{\mathbf{x}}^{q}_{0}).
\end{align}
Here, $\mathbf{a}_{p+q}$ represents the action words that combines the $p$ part and $q$ part. 
$\tau_3$ adjusts the strength of inference guidance. 
$\mathbf{H}^{-1}$ is the approximate inverse Hessian matrix used in the L-BFGS optimizer~\cite{liu1989limited}.
The function $G_m$ is used to measure the motion similarity in their global joint space, defined as:
\begin{align}
\| R(\text{slice}(\hat{\mathbf{x}}^{Inv}_{0}, p)) - R(\hat{\mathbf{x}}^{p}_{0}) \|_2^2 \notag  + \| R(\text{slice}(\hat{\mathbf{x}}^{Inv}_{0}, q)) - R(\hat{\mathbf{x}}^{q}_{0}) \|_2^2, \notag
\end{align}
where $\text{slice}(\mathbf{x}; i)$ represents the extraction of the segment $i$ from the motion $\mathbf{x}$, and $\hat{\mathbf{x}}^{Inv}_{0}$ is obtained similarly as in Eq. 8 in the main paper.

We take the final denoising step of $\hat{\mathbf{x}}^{Inv}$ as the result of motion blending between adjacent motions.
We iteratively run the above blending method on all the motion clips within a storyboard composing them into a complete animation.

\begin{table}[!tb]
\centering
\caption{Quantitative comparison of motion blending methods on the HumanML3D subset. }
\vspace{-2mm}
\label{tab:blending_results}
\renewcommand{\arraystretch}{1.2}
\resizebox{0.7\linewidth}{!}{
    \begin{tabular}{lccc}
    \toprule
    {Method} & {FID $\downarrow$} & {AUJ $\downarrow$} & {PJ (GT = 0.0330)} \\
    \midrule
    SQUAD       & 1.0015 & 0.5589 & 0.0002 \\
    DoubleTake  &\textbf{0.3156} & 0.2823 & 0.0182 \\
    Ours        & 0.4204 & \textbf{0.2075} & \textbf{0.0217} \\
    \bottomrule
    \end{tabular}
}
\end{table}

\paragraph{Performance evaluation.}
To evaluate its effectiveness, we conducted a quantitative comparison against a diffusion-based baseline DoubleTake~\cite{shafir2023human} and SQUAD interpolation. 
The evaluation was performed on a curated subset of the HumanML3D dataset, comprising 1,219 samples (25\% of the test set). This subset was selected by filtering for text descriptions containing two distinct action words, indicating the presence of multiple motion phases.
For each sample, we identified two key motion segments and masked out the \textbf{middle 20 frames} between the corresponding representative keyposes to define the transition region.
Each method was then tasked with generating the blended motion.
The resulting motions were compared against ground-truth transitions using the following widely adopted metrics:
\begin{itemize}
    \item \textbf{FID}: Fréchet Inception Distance, for assessing motion realism.
    \item \textbf{AUJ}: Area Under the Jerk Curve, measuring average motion smoothness.
    \item \textbf{PJ}: Peak Jerk, capturing extreme motion fluctuations. The mean PJ value of the ground truth samples in the test set is $0.0330$. The metric close to this value is better.
\end{itemize}
The evaluation results are reported in Tab.~\ref{tab:blending_results}.
Quantitative results indicate that our method surpasses SQUAD in both realism and smoothness and closely matches the performance of DoubleTake across all metrics.

}

\begin{figure*}[!tb]
    \centering
    \begin{overpic}[width=\textwidth]{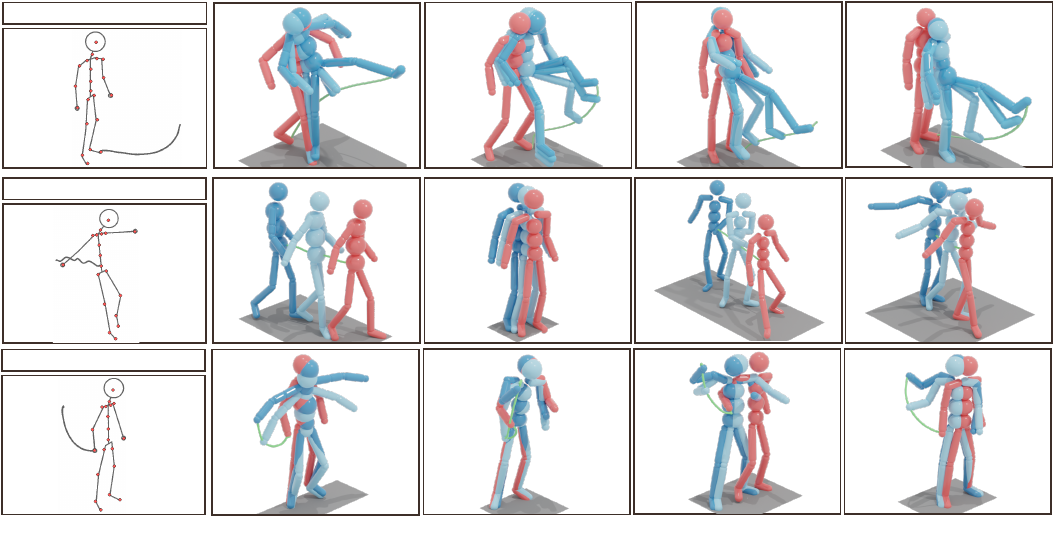} %
        \put(1,48.8) {\small Action: Kick}
        \put(1,32.2) {\small Action: Walk}
        \put(1,16.2) {\small Action: Raise}
        \put(8.2,0) {\small Input}
        \put(25,0) {\small Motion Retrieval}
        \put(45,0) {\small Lift-and-Control}
        \put(63.8,0) {\small Direct 2D-to-Motion}
        \put(89,0) {\small Ours}
    \end{overpic}
    \vspace{-6mm}
    \caption{\textbf{More Visual comparison.} Check either the keypose (red) or the trajectory (green) for the differences.
    }
    \vspace{-3mm}
    \label{fig:comp_supple}
\end{figure*}

\section{More Implementation Details}
\paragraph{Model Details}
Our pre-trained motion diffusion model is based on MLD~\cite{chen2023executing}. Both the trajectory ControlNet and the keypose adapter are composed of four Transformer encoder blocks.
For the text input, we first combine the action word $W_{a}$ with the template \emph{`a person [action\_word]'} to form a simplified text description. We then leverage a CLIP model~\cite{radford2021learning} to encode the text into embeddings $\va$ and use linear layers to project the timestep $t$ into time embeddings. These text embeddings are added to the time embeddings and concatenated with the noisy latent $\vz_t$.
The trajectory encoders (\ie $\mathcal{E}_{tr}^{3D}$ and $\mathcal{E}_{tr}^{2D}$), and and keypose encoders (\ie $\mathcal{E}_{k}^{3D}$ and $\mathcal{E}_{k}^{2D}$), as illustrated in Figs. 3 and 4 in the main paper, primarily consist of a single Transformer encoder designed to encode the trajectory and keypose into their corresponding embeddings.

\paragraph{Evaluation Details}
We follow the evaluation protocol for the \textit{Cross} setting from OmniControl~\cite{xie2024omnicontrol}, assessing multiple combinations of joints. A total of 63 combinations are randomly sampled during the evaluation process.

\paragraph{Inference Time}
To evaluate the inference efficiency of our approach and baseline methods, we report the average inference time of generating a single motion clip, measured in seconds, on a single NVIDIA RTX4090 GPU.
Specifically, our \name takes $0.427s$ to generate a motion clip, while the running time is $0.0415s$, $0.288s$, and $0.427s$ for Motion Retrieval, Lift-and-Control, Direct 2D-to-Motion methods, respectively. 
Since inference guidance is a post-optimization strategy, it takes around $0.058s$. The Direct 2D-to-Mition method employs the same inference pipeline as ours.

\section{Design Details of Competitors}
Since no existing framework directly addresses the problem of translating sketch storyboards into 3D animation, we developed three baseline approaches based on the latest works.
To ensure a fair comparison, all baseline methods use the same input and are trained and evaluated on the same dataset.
Additionally, \textit{Lift-and-Control} and \textit{Direct 2D-to-Motion} primarily adopt the same model architecture as \name, with only key components replaced. Fig.~\ref{fig:comp_supple} shows more visual comparison examples.

\paragraph{Motion Retrieval}
This retrieval baseline is primarily based on TMR~\cite{petrovich2023tmr}, retaining the same text encoder, motion encoder, and motion decoder.
We use the same 2D trajectory and keypose encoder as \name to extract trajectory and keypose embeddings, which are then concatenated with the text embedding.
The TEMOS~\cite{petrovich2022temos} losses and the InfoNCE loss~\cite{oord2018representation} are employed to train this baseline using the AdamW optimizer~\cite{loshchilov2017decoupled} with a learning rate of $10^{-4}$ and a batch size of 32.

\paragraph{Lift-and-Control}
This baseline first lifts the 2D keyposes and trajectories to 3D, then employs the same multi-conditional motion diffusion model as \name for motion synthesis.
The lifting component is based on MotionBERT~\cite{zhu2023motionbert}.
We separately project the 2D keyposes and trajectories into high-dimensional features and incorporate learnable spatial and temporal encodings. Subsequently, we apply the DSTformer module, as described in MotionBERT, to lift the keyposes and trajectories to 3D.
For training, we adopt the pretraining loss, with the only modification being the separation of the 2D re-projection loss into keypose and trajectory components.
The lifting network is trained for 1000 epochs with a learning rate of $5 \times 10^{-4}$ and a batch size of 64 using the AdamW optimizer~\cite{loshchilov2017decoupled}.

\paragraph{Direct 2D-to-Motion}
This baseline differs from \name by directly leveraging 2D keyposes and trajectories to train the multi-conditional motion diffusion model. The training strategy remains unchanged, using the same number of epochs and loss function, but replacing the 3D input with 2D input, as detailed in Section 4.3.

\section{More Discussion}

\begin{figure}[!h]
    \centering
    \includegraphics[width=\linewidth]{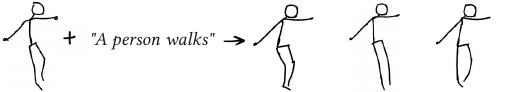}
    \vspace{-6mm}
    \caption{\textbf{Lifting 2D sequence}. Given a sketch keypose and the action word from a storyboard (left), we exploit the SoTA method \cite{gal2024breathing} to produce the animated animation sequence (right).
    }
    \vspace{-3mm}
    \label{fig:lifting_2d_sequence}
\end{figure}

\paragraph{Lifting 2D Animation Sequence.}
Given a storyboard frame, an alternative solution is to first generate a 2D animation sequence and then elevate it to its corresponding 3D motion.
There are two possible ways to generate a complete 2D animation sequence:
\begin{itemize}
    \item The first solution involves using sketch image animation methods~\cite{gal2024breathing,rai2024enhancing} to generate a sketch image animation sequence, estimating the 2D poses, and then lifting them to 3D (\eg using Sketch2Pose~\cite{brodt2022sketch2pose}). 
    However, as shown in Fig.\ref{fig:lifting_2d_sequence}, our experiments reveal that existing sketch image animation methods~\cite{gal2024breathing} fail to produce structurally consistent human sequences due to the lack of human body priors. Extending these methods to incorporate human body priors, such as SMPL or human skeletons, falls outside the scope of this paper.

    \item The second solution is to build a 2D motion diffusion model, similar to~\cite{kapon2024mas,li2024lifting}, conditioned on 2D keyposes and joint trajectories. After generating the 2D motion from the 2D keyposes and trajectories, the results can then be lifted to the full 3D motion. We did not evaluate this solution and leave it for future work.
\end{itemize}

\paragraph{Speed Control.}
\rev{
In future work, we have discussed the potential solution to handle over-sketched speed lines for motion speed control. 
A more straightforward way to achieve speed control in our method is non-uniform trajectory sampling. 
Given a fixed motion frame rate, a faster trajectory means fewer sampled points, covering a longer trajectory in a shorter time (fewer motion frames), and vice versa. 
In our interface, we can additionally record the user's drawing speed, sample trajectory points accordingly (instead of uniform sampling), and sequentially assign these points to corresponding motion frames. 
In this way, the resulting motion has the desired speed control as the drawing speed.
}

\begin{figure}[!t]
    \centering
    \begin{overpic}[width=0.9\linewidth]{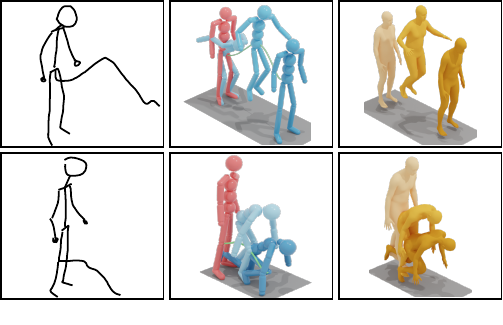}
        \put(6,1) {\footnotesize (a) Sketch frame}
        \put(34,1) {\footnotesize (b) Capsule-bone model}
        \put(72.5,1) {\footnotesize (c) SMPL model}
    \end{overpic}
    \vspace{-3mm}
    \caption{\textbf{Motion Visualization with different human models}. 
    }
    \vspace{-3mm}
    \label{fig:vis_smpl}
\end{figure}

\paragraph{SMPL model visualization}
\rev{
As stated in Sec. 3 in the main paper, we use a capsule-bone human model in our visualization. However, other models, \eg, the SMPL human model, are also feasible for visualization purposes. Fig. \ref{fig:vis_smpl} displays two examples visualized by both the capsule-hone human and the SMPL human, demonstrating consistent and high-quality resulting motion.
}

\section{User \rev{Evaluation} Details}
We conduct user studies using pairwise comparisons. In each test, participants watched two motions, generated by our model and a competitor, given the same sketch keypose, trajectory, and action word. 
Participants were asked to choose their preferred motion based on three evaluation criteria. 
A total of 21 pairs were tested, evenly divided into three groups for comparison with three competitors.
We invited 58 participants to complete the survey, which was conducted online using Google Forms.

The three key criteria are motion realism, keypose accuracy, and trajectory accuracy, ensuring a comprehensive assessment of the motion quality. 
For motion realism, participants are asked to assess how natural and realistic the generated motion appeared, focusing on the smoothness and fluidity of movements. 
For keypose accuracy, they evaluate how closely the generated animation matched the keypose provided in the input sketch. The corresponding keypose in the motion sequence is highlighted in red as in the paper and video.
For trajectory accuracy, participants are asked to evaluate how well the animation followed the input joint trajectory, observing whether the motion adhered to the expected direction and maintained a consistent flow. 
Each animation pair was presented side by side, allowing participants to compare the results objectively. For a given input, the participants should indicate their preference on all three criteria.

\end{document}